\documentclass{jfm}
\usepackage{amsmath}
\usepackage{amssymb}
\usepackage{graphicx}
\usepackage{epstopdf, epsfig}
\usepackage{psfrag}
\usepackage{subcaption}
\usepackage{cleveref}
\usepackage{xcolor}
\usepackage{mathrsfs}

\newtheorem{innercustomgeneric}{\customgenericname}
\providecommand{\customgenericname}{}
\newcommand{\newcustomtheorem}[2]{%
  \newenvironment{#1}[1]
  {%
   \renewcommand\customgenericname{#2}%
   \renewcommand\theinnercustomgeneric{##1}%
   \innercustomgeneric
  }
  {\endinnercustomgeneric}
}

\newcustomtheorem{customthm}{Theorem}
\newcustomtheorem{customlemma}{Inequality}

\newcommand\bcolon{\boldsymbol{\colon}}
\newcommand{\bs}[1]{\boldsymbol{#1}}
\newcommand*{\QEDB}{\hfill\ensuremath{\square}}

\shorttitle{Pressure-driven flows in helical pipes}

\shortauthor{A. Kumar}

\title{Pressure-driven flows in helical pipes: bounds on flow rate and friction factor}

\author{Anuj Kumar\aff{1}
  \corresp{\email{akumar43@ucsc.edu}}}

\affiliation{\aff{1}Department of Applied Mathematics, Baskin School of Engineering,  University of California, Santa Cruz, CA 95064, USA}

\begin{document}

\maketitle

\begin{abstract}
In this paper, we use the well-known background method to obtain a rigorous lower bound on the volume flow rate through a helical pipe driven by a pressure differential in the limit of large Reynolds number. As a consequence, we also obtain an equivalent upper bound on the friction factor. These bounds are also valid for toroidal and straight pipes as limiting cases. By considering a  two-dimensional background flow with varying boundary layer thickness along the circumference of the pipe, we obtain these bounds as a function of the curvature and torsion of the pipe and therefore capture the geometrical aspects of the problem.  In this paper, we also present a sufficient criterion to find out which pressure-driven flow and surface-velocity-driven flow problems can be tackled using the background method.
\end{abstract}


\section{Introduction}
\label{Introduction}
Curved pipes have a wide range of applications in the industry because of their enhanced mixing properties, high heat transfer coefficient, and compact structure. Examples of application include, but are not limited to, heat exchangers, air-conditioning systems, chemical reactors, and steam generators \citep[see the review by][]{vashisth2008review, naphon2006review}. One of the crucial questions in the study of turbulent flows in curved pipes is the accurate determination of the dependence of the flow rate and friction factor on the applied pressure difference between the two ends of the pipe, and its dependence on geometrical parameters such as the pipe's curvature and torsion. The extensive usage of curved pipes in the industry has motivated many studies to characterize this dependence \citep[see][]{ito1959friction, liu1993axially, yamamoto1994torsion, yamamoto1995experimental, cioncolini2006experimental}. However, only  a few of these studies consider the high Reynolds number limit, which is the objective of this paper.

The flow structure inside a curved pipe can vary substantially with Reynolds number and pipe geometry, which leads to a number of different regimes, each with its own distinct functional dependence of the flow rate and friction factor on these parameters. As such, quantifying this dependence becomes difficult even for the laminar flow, unlike the case of a straight pipe. Indeed, at low Reynolds number, an imbalance between centrifugal force and cross-stream pressure leads to the onset of secondary counter-rotating vortices known as Dean's vortices, which were first experimentally observed by \citet{eustice1910flow, eustice1911experiments}. \citet{dean1927xvi, dean1928lxxii} confirmed this observation analytically in the low curvature limit by computing the flow velocity as a perturbation of the well-known laminar Poiseuille flow solution. \citet{dean1928lxxii} showed that the effect of curvature is to decrease the flow rate and that this effect is of second-order, i.e. quadratic in curvature. Several other studies were performed in the limit of small curvature to obtain a steady-state flow solution in a toroidal pipe, see for example, \citet{mcconalogue1968motion, van1978extended, dennis1980calculation}. For a comprehensive review of the topic, the reader is referred to \citet{berger1983flow}. \citet{germano1982effect} further extended Dean's result to a helical pipe with small torsion and \citet{tuttle1990laminar} showed that small torsion leads to a second-order decrease in the flow rate. However, no analytical result exists for the steady flow in a pipe with a finite radius of curvature or torsion. Therefore, even in the laminar regime, one has to rely on empirical formulae to quantify the flow rate.

The transition to turbulence in curved pipes also differs substantially from the case of a straight pipe. \citet{taylor1929criterion} and \citet{white1929streamline} found that flow in a curved pipe is more stable than in a straight pipe. Notably, they saw that the critical Reynolds number for the transition is twice as large as in the straight pipe case. Inspired by this observation, \citet{sreenivasan1983stabilization} conducted experiments in a straight tube followed by a helical tube with curvature $\kappa = 0.058$.  They noticed an oscillating behavior near the inner wall of the helical tube at a moderate Reynolds number,  which \citet{webster1993experimental, webster1997traveling} attributed to the presence of traveling wave perturbations to the Dean's vortices. Recent years have witnessed a resurgence in carefully conducted studies to quantify the effect of curvature on the stability of flow in a torus. \citet{kuhnen2015subcritical} studied this problem using a novel experimental setup where a magnetically controlled steel sphere drives the flow in a torus. They conjectured that the transition switches from subcritical to supercritical for a critical torus curvature $\kappa \simeq 0.028$. Soon after that, \citet{canton2016modal} performed an in-depth linear stability analysis, covering the entire curvature range, and obtained the critical Reynolds number as a function of the curvature. More recently, \citet{canton2020critical} have shed light on the complexity of transition for flow in a torus, demonstrating in particular that for $\kappa \simeq 0.025$, two branches of solution can coexist at the same Reynolds number: one with subcritically-excited sustained turbulence, and the other consisting of  a low-amplitude travelling wave originating from a supercritical Hopf bifurcation. 

The incredible complexity of curved pipe flows makes it impossible to obtain the precise dependence of mean quantities such as flow rate or friction factor on model parameters. This is especially true at high Reynolds number, where both laboratory experiments and numerical computations are extremely challenging and must be repeated for different pipe geometries. As noted by \citet{kalpakli2016turbulent}, the determination of the friction factor (or equivalently the flow rate) for turbulent flows in curved pipes has generally been neglected, with only a few exceptions \citep{ito1959friction, cioncolini2006experimental}.  However, as we shall demonstrate in this paper, it is possible to obtain bounds on these mean quantities as explicit functions of flow and geometric variables, in the high Reynolds number limit.

Obtaining bounds on mean quantities in fluid mechanics goes back to the classical technique of \citet{howard1963heat}, which was further developed by \citet{busse1969howard, busse1970bounds}.  In the 1990s, Doering and Constantin \citep{PhysRevLett.69.1648, PhysRevE.49.4087, PhysRevE.51.3192, PhysRevE.53.5957}, based on the ideas from \citet{hopf1955lecture}, developed a new technique known as the background method to bound mean quantities. This method requires a careful choice of a trial function (the background field) to satisfy a spectral constraint in order to obtain a bound on the desired quantity. Since the work of Doering and Constantin, this method has been applied to a wide variety of problems in fluid dynamics. Examples include  upper bounds on the rate of energy dissipation in surface-velocity-driven flows \citep{PhysRevLett.69.1648, PhysRevE.49.4087, marchioro1994remark, wang1997time,  plasting2003improved}, pressure-driven flows \citep{PhysRevE.51.3192}, and surface-stress-driven flows \citep{tang2004bounds, hagstrom2014bounds};  upper bounds on the heat transfer in different configurations of Rayleigh--B\'enard convection \citep{PhysRevE.53.5957, doering2001upper, otero2002bounds, plasting2005infinite, wittenberg2010bounds, whitehead2011ultimate, whitehead2014rigorous, goluskin2015internally, goluskin2016bounds, fantuzzi2018boundsB} and B\'enard--Marangoni convection \citep{hagstrom2010bounds, fantuzzi2018boundsA, fantuzzi2020new}; upper bounds on buoyancy flux in stably stratified shear flows \citep{caulfield2001maximal, caulfield2005buoyancy}.

In this paper, we use this background method to obtain  a lower bound on the flow rate and an equivalent upper bound on friction factor for flows in helical pipes.  The novelty in this paper is the use of a two-dimensional background flow in contrast with most previous applications of the background method, where the geometry was simple enough to use a one-dimensional background flow to suffice the desired purpose. We start by setting up the problem in \S \ref{Problem Setup}, where we describe the flow configuration and the coordinate system used to solve the problem. In \S \ref{The background method formulation}, we formulate the background method in the context of pressure-driven flows in helical pipes. In \S \ref{Lower bound on the volume flow rate}, we choose the background flow and obtain bounds on the flow rate and friction factor. Finally, in \S \ref{Discussion and concluding remarks}, we compare our findings with  available experimental data and make a few remarks about the applicability of the background method to other interesting problems in engineering.

\section{Problem Setup}
\label{Problem Setup}
\subsection{Flow configuration}
\label{Flow configuration}
\begin{figure}
\centering
\begin{tabular}{lc}
\begin{subfigure}{0.5\textwidth}
\centering
 \includegraphics[scale = 1.75]{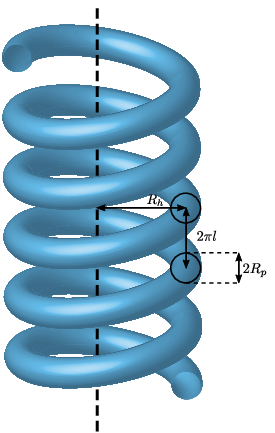}
\caption{}
\end{subfigure} &
\begin{subfigure}{0.5\textwidth}
\centering
 \includegraphics[scale = 0.3]{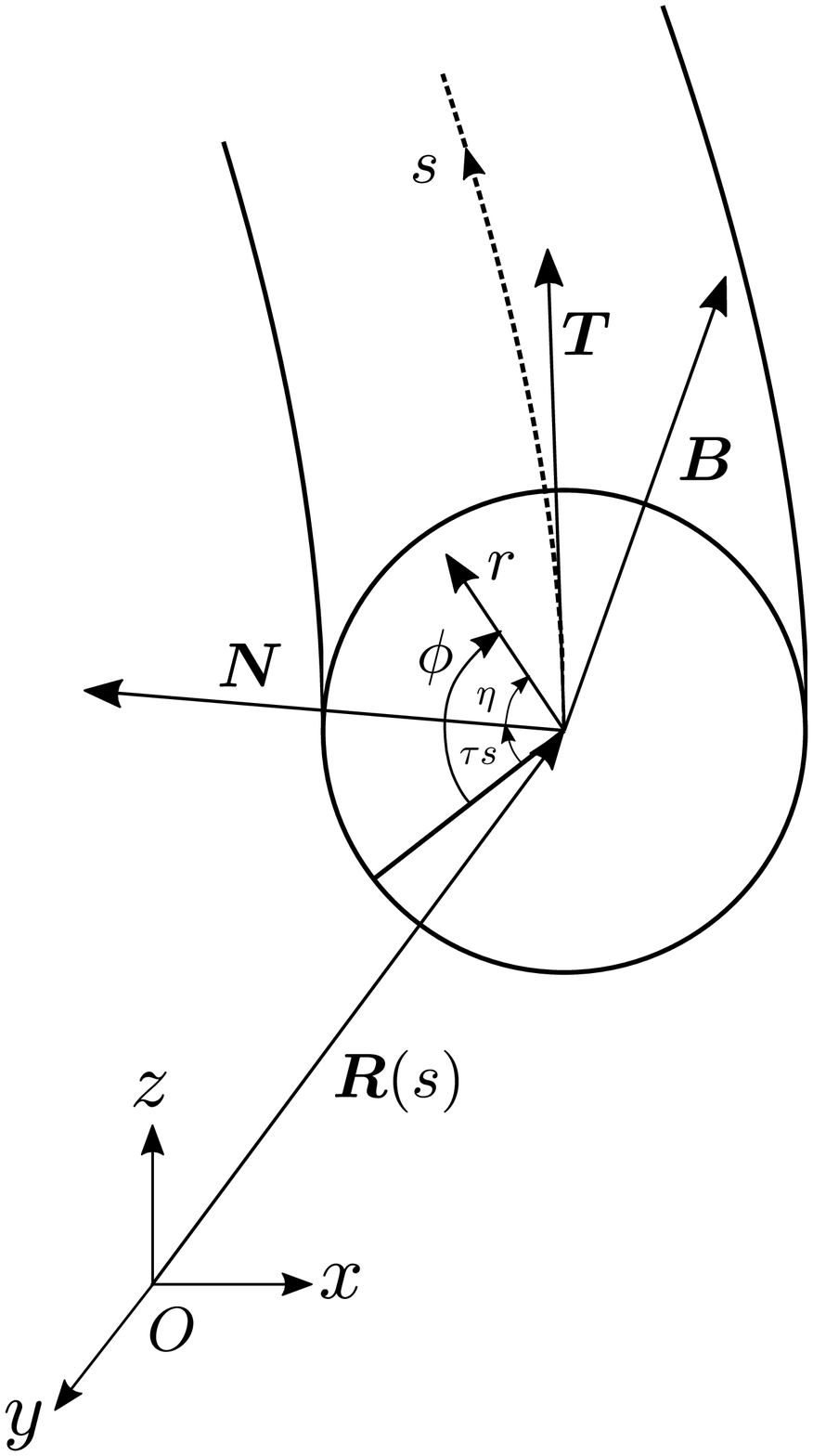}
\caption{}
\end{subfigure}
\end{tabular}
 \caption{(a) Schematic diagram of a helical pipe with radius $R_p$, radius of the centerline helix $R_h$, and pitch of the centerline helix $2 \upi l$.  The dashed line is the axis of rotation of the helical pipe. (b) Illustration of the coordinate system $(s, r, \phi)$ used in this paper.}
 \label{Problem schematic}
\end{figure}

We consider the flow of an incompressible fluid with density $\rho$ and kinematic viscosity $\nu$ in a helical pipe. The radius of the pipe is denoted as  $R_p$, the radius of the centerline helix $R_h$, and the pitch of the centerline helix is $2 \upi l$ (see figure \ref{Problem schematic}a). Here, the centerline helix refers to the locus of the center of the pipe. The flow is driven by a body force $\bs{f}^\ast$, which has dimensional amplitude $F$. The choice of forcing is described in \S \ref{choice of forcing}. We non-dimensionalize the variables as follows
\begin{eqnarray}
\bs{f} = \frac{\bs{f}^\ast}{F}, \quad \bs{u} = \left(\frac{\rho}{F R_p}\right)^{\frac{1}{2}}\bs{u}^\ast,  \quad p = \frac{p^\ast - p_a}{F R_p}, \quad \bs{x} = \frac{\bs{x}^\ast}{R_p}, \quad t = \left(\frac{F}{\rho R_p}\right)^{\frac{1}{2}} t^\ast.
\label{Flow configuration: nondim}
\end{eqnarray}
Here, $p_a$ is the ambient pressure, whereas $\bs{f}$, $\bs{u}$, $p$, $\bs{x}$, and $t$ denote the non-dimensional forcing, velocity, pressure, position, and time, respectively. Quantities with a star in superscript are dimensional.  The equations governing the flow in non-dimensional form are as follows
\begin{eqnarray}
&& \bnabla \bcdot \boldsymbol{u} = 0, \nonumber \\
&& \frac{\partial \boldsymbol{u}}{\partial t} + \boldsymbol{u} \bcdot \bnabla \boldsymbol{u} = - \bnabla p + \frac{1}{\Rey} \nabla^2 \boldsymbol{u} + \boldsymbol{f},
\label{Flow configuration: governing equations}
\end{eqnarray}
where $$\Rey =  \frac{R_p}{\nu} \left(\frac{F R_p}{\rho}\right)^{\frac{1}{2}}$$ is the Reynolds number. The boundary conditions at the surface of the pipe are no-slip and impermeable.

\subsection{Coordinate system}
\label{Coordinate system}
In this subsection, we construct an orthogonal coordinate system that is well-suited for our problem. This coordinate system was first introduced by \citet{germano1982effect}, who was interested in the effect of small torsion on Dean's solution. The coordinate system has been extensively used since then in both analytical and computational studies of flows in helical pipes \citep{kao1987torsion, germano1989dean, tuttle1990laminar, liu1993axially, yamamoto1994torsion, huttl2001direct, gammack2001flow}. For clarity and self-consistency, we repeat its construction below.

Let $a$ and $2 \upi b$ be the non-dimensional centerline helix radius and pitch, where $a = R_h / R_p$ and $b = l / R_p$. The equation of this helix parameterized with arc length $s$ in a Cartesian coordinate system $(x, y, z)$ is given by
\begin{eqnarray}
(x(s), y(s), z(s)) = \left(a \cos\left(\frac{s}{\sqrt{a^2 + b^2}}\right), a \sin\left(\frac{s}{\sqrt{a^2 + b^2}}\right), \frac{b s}{\sqrt{a^2 + b^2}}\right).
\label{Coordinate system: center line helix}
\end{eqnarray}
Let $\bs{R}(s) = (x(s), y(s), z(s))$ be the position vector and let $\bs{T}(s)$, $\bs{N}(s)$, and $\bs{B}(s)$ be the tangent, normal, and binormal to the centerline helix, which are given by
\begin{eqnarray}
\bs{T} = \frac{d \bs{R}}{ds}, \quad \bs{N} = \frac{1}{\kappa} \frac{d \bs{T}}{d s}, \quad \bs{B} = \bs{T} \times \bs{N}.
\label{Coordinate system: def TNB}
\end{eqnarray}
The relations among the tangent, normal, and binormal are given by the Frenet--Serret formulae, which are
\begin{eqnarray}
\frac{d \bs{N}}{d s} = \tau \bs{B} - \kappa \bs{T}, \quad \frac{d \bs{B}}{d s} = - \tau \bs{N},
\label{Coordinate system: Frenet--Serret formulae}
\end{eqnarray}
where 
\begin{eqnarray}
\kappa = \frac{a}{a^2 + b^2}, \quad \text{and} \quad \tau = \frac{b}{a^2 + b^2}
\label{Coordinate system: def kappa tau}
\end{eqnarray}
are the non-dimensional curvature and torsion of the helix. The curvature is considered smaller than one ($\kappa < 1$) in this paper. We now construct a coordinate system $(s, r, \eta)$ such that any Cartesian position vector $\bs{x}$ can be expressed as
\begin{eqnarray}
\bs{x} = \bs{R} + r \cos \eta \bs{N} + r \sin \eta \bs{B}.
\label{Coordinate system: s r eta}
\end{eqnarray}
With the use of (\ref{Coordinate system: def TNB}) and (\ref{Coordinate system: Frenet--Serret formulae}),
\begin{eqnarray}
d \bs{x} \bcdot d \bs{x} = \left[(1 - r \kappa \cos \eta)^2 + \tau^2 r^2\right] ds^2 + dr^2 + r^2 d\eta^2 + 2 \tau r^2 ds d\eta.
\label{Coordinate system: line element s r eta}
\end{eqnarray}
Therefore, the resulting coordinate system is non-orthogonal. However, using the transformation $\eta = \phi - \tau s$ in (\ref{Coordinate system: line element s r eta}), we obtain
\begin{eqnarray}
d \bs{x} \bcdot d \bs{x} = (1 - r \kappa \cos(\phi - \tau s))^2 ds^2 + dr^2 + r^2 d\phi^2.
\label{Coordinate system: line element s r phi}
\end{eqnarray}
The coordinate system $(s, r, \phi)$ is orthogonal, and will be used to perform calculations in the rest of the paper. The scale factors for this coordinate system are defined as
\begin{eqnarray}
h_s = (1 - r \kappa \cos(\phi - \tau s)), \quad h_r = 1, \quad h_{\phi} = r.
\label{Coordinate system: line element s r phi scale factors}
\end{eqnarray}

The impermeability and no-slip condition at the surface of the pipe in the $(s, r, \phi)$ coordinate system translate to
\begin{eqnarray}
\bs{u} = (u_s, u_r, u_{\phi}) = \bs{0} \text{ at } r = 1.
\label{Coordinate system: boundary condition}
\end{eqnarray}
In this paper,  we assume that the flow is periodic in the streamwise direction $s$ with period $s_p$. Hence, the domain of interest in the $(s, r, \phi)$ coordinate system is
\begin{eqnarray}
\Omega = [0, s_p]\times[0, 1]\times[0, 2\upi].
\label{Coordinate system: domain of interest}
\end{eqnarray}

\subsection{Choice of forcing}
\label{choice of forcing}
We choose to drive the flow with a  dimensional forcing
\begin{eqnarray}
\bs{f}^\ast = -\frac{1}{1 - r^\ast \kappa^\ast \cos(\phi^\ast - \tau^\ast s^\ast)} \times \frac{d P}{d s^\ast} \; \bs{e}_{s} \quad \text{for } 0 \leq r^\ast \leq R_p. 
\label{Coordinate system: choice of forcing}
\end{eqnarray}
Here, $-d P/d s^\ast$ is a constant and can be thought of as the applied pressure gradient. Note how this streamwise directed forcing varies across the cross-section. The reason for this choice of forcing over a conventional forcing, which would be constant across the cross-section, is that the line integral along the streamwise direction for this forcing is independent of the position on the pipe cross-section and depends only on the difference of streamwise coordinates, i.e.
\begin{eqnarray}
\int_{s^\ast = s_1^\ast}^{s_2^\ast} \bs{f}^\ast \bcdot \bs{e}_{s} \; dl = \int_{s^\ast = s_1^\ast}^{s_2^\ast} - \frac{d P}{d s^\ast} \; ds^\ast =  -\frac{d P}{d s^\ast} (s_2^\ast - s_1^\ast),
\end{eqnarray}
where $d l = h^\ast_s ds^\ast$ is the line element with $h^\ast_s = 1 - r^\ast \kappa^\ast \cos(\phi^\ast - \tau^\ast s^\ast)$. By contrast, for the conventional forcing, the value of this line integral would also depend on the position on the cross-section. Hence, we believe that our choice of forcing is good for modeling a flow driven by constant pressure boundary conditions. More detail on this choice of forcing in the context of flow in a torus can be found in \citet{canton2016modal, canton2017characterisation, rinaldi2019vanishing}. Note that in the limit of vanishing curvature ($\kappa \to 0$), our choice does reduce to constant forcing in the streamwise direction and therefore is consistent with the usual modeling of pressure-driven flow in a straight pipe. Based on (\ref{Coordinate system: choice of forcing}), we define the forcing scale as $$F = -\frac{d P}{d s^\ast}.$$
This implies that the non-dimensional forcing is given by
\begin{eqnarray}
\bs{f} = \frac{1}{1 - r \kappa \cos(\phi - \tau s)} \bs{e}_{s} \quad \text{for } 0 \leq r \leq 1.
\label{Coordinate system: nondim forcing}
\end{eqnarray}

\subsection{Quantities of interest}
\label{Quantities of interest}
We are interested in obtaining a lower bound on the average non-dimensional flow rate $Q$, which we simply call flow rate, and an equivalent upper bound on the friction factor $\lambda$ in the limit of high Reynolds number. As we are concerned with the high Reynolds number limit, so we use an inertial scaling to define the non-dimensional flow rate $Q$ as
\begin{eqnarray}
Q = \frac{1}{R_p^2} \left(\frac{\rho}{F R_p}\right)^{\frac{1}{2}} Q^\ast = \left\langle  \int_{\phi = 0}^{2 \upi} \int_{r = 0}^{1}  u_s r dr d\phi  \right\rangle,
\label{Coordinate system: nondim volume flow rate}
\end{eqnarray}
where $Q^\ast$ is the long-time average of the dimensional flow rate, $u_s$ is the streamwise component of the non-dimensional velocity field $\bs{u}$ and
\begin{eqnarray}
\langle [\; \cdot \;] \rangle  = \lim_{T \to \infty} \frac{1}{T} \int_{t = 0}^{T} [\; \cdot \;] \; dt
\label{The background method formulation: def time-avg}
\end{eqnarray}
denotes the long-time average of a quantity. The Darcy--Weisbach friction factor $\lambda$, which is four times the Fanning friction factor, is defined as
\begin{eqnarray}
\lambda = - \frac{d P}{d s^\ast} \; \frac{4 R_p}{\rho u_m^{\ast 2}},
\end{eqnarray}
where $u_m^\ast$ is the dimensional streamwise mean velocity given by
\begin{eqnarray}
u_m^\ast = \frac{Q^\ast}{\pi R_p^2}.
\end{eqnarray}
When expressed in non-dimensional variables, the friction factor is
\begin{eqnarray}
\lambda = \frac{4 \pi^2}{Q^2}.
\label{The background method formulation: fric frac}
\end{eqnarray}
From (\ref{The background method formulation: fric frac}), we notice that a lower bound on the flow rate $Q$ will provide an upper bound on the friction factor $\lambda$.

\section{The background method formulation}
\label{The background method formulation}
In this section, we describe the general approach of the background method applied to our problem. The formulation that we develop here is for any general background flow field and is similar to the one given in \citet{PhysRevE.51.3192} for pressure-driven channel flow. 

The background method, in essence, works as follows. We first derive time-averaged integral identities from the governing equations (\ref{Flow configuration: governing equations}) (using the fact that the long-time averages of certain time derivatives vanish) in order to rewrite the quantity of interest $Q$ given by (\ref{Coordinate system: nondim volume flow rate}) as an equivalent long-time averaged expression that is easier to bound using analysis techniques. To that end, we begin by establishing a time-averaged total energy equation,  by taking the dot product of equation (\ref{Flow configuration: governing equations}) with $\boldsymbol{u}$ and then by performing a volume integration on the resulting equation. The result is
\begin{eqnarray}
\frac{1}{2}\frac{d ||\bs{u}||^2_2}{d t}  = - \frac{1}{\Rey} || \bnabla \bs{u}||^2_2 + \int_{\Omega} \bs{f} \bcdot \bs{u} \; d \bs{x},
\label{The background method formulation: total energy equation}
\end{eqnarray}
where $||\cdot||_2$ denotes the $L^2$-norm, which is given by
\begin{eqnarray}
||\cdot||_2 = \left(\int_{\Omega} |\cdot|^2 \; d \bs{x}\right)^{\frac{1}{2}},
\label{The background method formulation: L2}
\end{eqnarray}
and  where the volume integral in $(s, r, \phi)$ coordinates is written as
\begin{eqnarray}
\int_{\Omega} [\; \cdot \;] \; d \bs{x} = \int_{s = 0}^{s_p} \int_{\phi = 0}^{2 \upi} \int_{r = 0}^{1} [\; \cdot \;] \;  h_s h_r h_\phi dr d\phi ds.
\label{The background method formulation: volume integral}
\end{eqnarray}
 The quantity $||\bs{u}||_2^2(t)$ can be shown to be uniformly bounded in time within the framework of the background method \citep[see][for example]{PhysRevLett.69.1648, PhysRevE.51.3192}. Therefore, the long-time average of the time derivative of $||\bs{u}||_2^2(t)$ vanishes.
As a result, taking the long-time average of  equation (\ref{The background method formulation: total energy equation}) leads to
\begin{eqnarray}
\left\langle \int_{\Omega} \bs{f} \bcdot \bs{u} \; d \bs{x} \right\rangle = \frac{1}{\Rey} \langle ||\bnabla \bs{u}||^2_2  \rangle .
\label{The background method formulation: time-avg energy equation}
\end{eqnarray} 
The second step of the method is to perform the background decomposition. We start by  writing the total velocity  $\bs{u}$ as the sum of two divergence-free velocity fields $\bs{u} = \bs{U} + \bs{v}$, where $\bnabla \bcdot \boldsymbol{U} = 0$ and $\bnabla \bcdot \boldsymbol{v} = 0$. We call $\bs{U}$ the background flow, which is steady and satisfies the same boundary conditions as
the full flow $\bs{u}$, while the perturbation $\bs{v}$ satisfies the homogeneous version of the boundary conditions.
The equation governing the evolution of $\bs{v}$ is given by 
\begin{eqnarray}
\frac{\partial \bs{v}}{\partial t} + \bs{U} \bcdot \bnabla \bs{U} + \bs{U} \bcdot \bnabla \bs{v} + \bs{v} \bcdot \bnabla \bs{U} + \bs{v} \bcdot \bnabla \bs{v} 
 = -  \bnabla p + \frac{1}{\Rey} \nabla^2 \bs{U} + \frac{1}{\Rey} \nabla^2 \bs{v} + \bs{f}.
\label{The background method formulation: peturb eqn.}
\end{eqnarray}
Taking the dot product of the above equation with $\boldsymbol{v}$ and performing a volume integration, followed by taking the long-time average, results in
\begin{eqnarray}
\left\langle \int_{\Omega} \bs{f} \bcdot \bs{v} \; d \bs{x} \right\rangle =   \frac{1}{2 \Rey} \left\langle ||\bnabla \bs{u}||_2^2 \right\rangle -  \frac{1}{2 \Rey}  ||\bnabla \bs{U}||_2^2 + \left\langle \mathcal{H}(\bs{v}) \right\rangle , \quad &&
\label{The background method formulation: time-avg perturb energy}
\end{eqnarray} 
where 
\begin{eqnarray}
\mathcal{H}(\bs{v}) =  \underbrace{\int_{\Omega} (\bs{v} \bcdot \bnabla \bs{U}_{\text{sym}}) \bcdot \bs{v} \; d \bs{x}}_{I}  +  \underbrace{\int_{\Omega} (\bs{U} \bcdot \bnabla \bs{U}) \bcdot \bs{v} \; d \bs{x}}_{II} + \underbrace{\frac{1}{2 \Rey} ||\bnabla \bs{v}||_2^2}_{III},
\label{The background method formulation: def H(v)}
\end{eqnarray}
and  $\bnabla \bs{U}_{\text{sym}}$ is the symmetric part of $\bnabla \bs{U}$, i.e.
\begin{eqnarray}
\bnabla \bs{U}_{\text{sym}} = \frac{\bnabla \bs{U} + \bnabla \bs{U}^\intercal}{2}.
\label{The background method formulation: sym of grad U}
\end{eqnarray}
We have used the following identity in deriving the equation (\ref{The background method formulation: time-avg perturb energy})
\begin{eqnarray}
|\bnabla \bs{u}|^2 = |\bnabla \bs{U}|^2 + |\bnabla \bs{v}|^2 + 2 \bnabla \bs{U} \bcolon \bnabla \bs{v},
\label{The background method formulation: a useful identity}
\end{eqnarray}
where, in index notation,
\begin{eqnarray}
\bnabla \boldsymbol{U} \bcolon \bnabla \boldsymbol{v} = \partial_{i} v_j \partial_{i} U_j.
\end{eqnarray}
Multiplying (\ref{The background method formulation: time-avg perturb energy}) by two and subtracting (\ref{The background method formulation: time-avg energy equation}) yields
\begin{eqnarray}
2 \left\langle \int_{\Omega} \bs{f} \bcdot \bs{v} \; d \bs{x} \right\rangle - \left\langle \int_{\Omega} \bs{f} \bcdot \bs{u} \; d \bs{x} \right\rangle = - \frac{1}{ \Rey}  ||\bnabla \bs{U}||_2^2 + 2 \left\langle\mathcal{H}(\bs{v})\right\rangle.
\label{The background method formulation: time-avg perturb energy about to be useful}
\end{eqnarray}
The left-hand side of (\ref{The background method formulation: time-avg perturb energy about to be useful}) can be simplified as follows
\begin{eqnarray}
2 \left\langle \int_{\Omega} \bs{f} \bcdot \bs{v} \; d \bs{x} \right\rangle - \left\langle \int_{\Omega} \bs{f} \bcdot \bs{u} \; d \bs{x} \right\rangle  = \left\langle \int_{\Omega} \bs{f} \bcdot \bs{u} \; d \bs{x} \right\rangle - 2 \left\langle \int_{\Omega} \bs{f} \bcdot \bs{U} \; d \bs{x} \right\rangle &&  \nonumber \\
= \left\langle \int_{s = 0}^{s_p} \left[ \int_{\phi = 0}^{2 \upi} \int_{r = 0}^{1} u_s r dr d\phi \right] ds \right\rangle - 2 \int_{s = 0}^{s_p} \left[ \int_{\phi = 0}^{2 \upi} \int_{r = 0}^{1} U_s r dr d\phi \right] ds \qquad && \nonumber \\
= s_p \left\langle \int_{\phi = 0}^{2 \upi} \int_{r = 0}^{1} u_s r dr d\phi \right\rangle - 2 s_p \int_{\phi = 0}^{2 \upi} \int_{r = 0}^{1} U_s r dr d\phi. \qquad \qquad \qquad \qquad &&
\label{The background method formulation: intermediate calc}
\end{eqnarray}
Note that we used $\bs{v} = \bs{u} - \bs{U}$ in the first line, then substituted the expression for $\bs{f}$ from (\ref{Coordinate system: nondim forcing}) and used the time independence of $\bs{U}$ to obtain the second line. The terms in the square brackets in the second line represent the flow of $\bs{u}$ and $\bs{U}$ through a cross-section of pipe and therefore are independent of the streamwise direction $s$ because of the  incompressibility of $\bs{u}$ and $\bs{U}$. Hence, we can easily integrate these expressions with respect to $s$, which leads to the third line. Using (\ref{The background method formulation: intermediate calc}) in (\ref{The background method formulation: time-avg perturb energy about to be useful}) and dividing by $s_p$ on both sides gives
\begin{eqnarray}
Q = \left\langle \int_{\phi = 0}^{2 \upi} \int_{r = 0}^{1} u_s r dr d\phi \right\rangle = 2 \int_{\phi = 0}^{2 \upi} \int_{r = 0}^{1} U_s r dr d\phi - \frac{1}{s_p \Rey}  ||\bnabla \bs{U}||_2^2 
+ \frac{2}{s_p} \left\langle\mathcal{H}(\bs{v})\right\rangle. \nonumber \\
\label{The background method formulation: time-avg perturb energy useful}
\end{eqnarray}
If one can prove that 
\begin{eqnarray}
\mathcal{H}(\bs{v}) + \gamma \geq 0 \quad \forall \; \bs{v}
\label{The background method formulation: spectral constraint}
\end{eqnarray}
for  a background flow $\bs{U}$ and some constant $\gamma$, then we have the following bound on the flow rate
\begin{eqnarray}
Q \geq 2 \int_{\phi = 0}^{2 \upi} \int_{r = 0}^{1} U_s r \; dr d\phi - \frac{1}{s_p \Rey}  ||\bnabla \bs{U}||_2^2 - \frac{2 \gamma}{s_p}.
\label{The background method formulation: bound on Q}
\end{eqnarray}
Following the convention \citep{PhysRevE.49.4087, PhysRevE.51.3192}, we call (\ref{The background method formulation: spectral constraint}) the spectral constraint. 

Note that the background method formulation given in \citet{PhysRevE.51.3192} for pressure-driven channel flows assumes that the background flow $\bs{U}$ is unidirectional and planar (a choice that is only suitable for planar geometries). As a result, the term $II$ in (\ref{The background method formulation: def H(v)}) is zero and therefore the functional $\mathcal{H}(\bs{v})$ is homogenous in their work. Here, we have given the background method formulation for a general background flow $\bs{U}$. Also, as we shall see, the choice of the background flow which works in the present case, is two-dimensional  which leads to a nonzero term $II$ in (\ref{The background method formulation: def H(v)}) and therefore the resultant functional $\mathcal{H}(\bs{v})$ is inhomogenous.

\section{Bounds on flow rate and friction factor}
\label{Lower bound on the volume flow rate}
In this section, we obtain a lower bound on the flow rate and an equivalent upper bound on the friction factor. We choose a family of background flows with varying boundary layer thickness along the circumference of the pipe. This variation of the boundary layer thickness will be carefully selected so that the spectral constraint (\ref{The background method formulation: spectral constraint}) is satisfied while optimizing the bound (\ref{The background method formulation: bound on Q})  simultaneously for different values of curvature $\kappa$ and torsion $\tau$, thereby obtaining a geometrical dependence on these parameters. Note that in this paper, the \textit{boundary layer} refers to the term boundary layer used in the context of the background method \citep[see for instance,][]{PhysRevE.49.4087, goluskin2016bounds} and is not the conventional viscous boundary layer. 
\subsection{Choice of background flow}
We make the following choice of background flow
\begin{figure}
\centering
 \includegraphics[scale = 0.5]{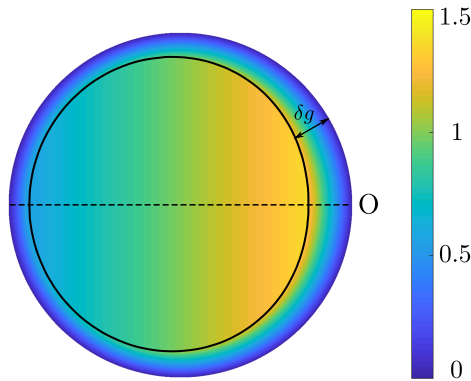}
 \caption{  Variation of the streamwise component $U_s$ of the background flow (\ref{Lower bound on the volume flow rate: the background flow}) across a cross-section of the pipe. In this example, the pipe's curvature is $\kappa = 0.5$ and torsion is $\tau = 0.25$. The solid black curve shows the edge of the boundary layer with variable thickness $\delta g(s, \phi)$. The point O denotes the outer edge of the pipe, i.e., the point on the cross-section, which is farthest from the axis of rotation of the helical pipe. The background flow in this figure corresponds to $\Lambda = 1$, and the boundary layer shape $g(s, \phi)$ is  given by (\ref{Lower bound on the volume flow rate: g c1 c2}), which is the shape obtained in the process of optimizing the bound.  }
\label{The background flow illustration}
\end{figure} 
\begin{eqnarray}
\bs{U}(s, r, \phi) = 
\begin{cases}
\left(\Lambda (1 - r \kappa \cos(\phi - \tau s)), \; 0, \; \Lambda \tau r \right) & \quad \text{if} \quad 0 \leq r < 1 - \delta g(s, \phi) \\
 \left(\Lambda (1 - r \kappa \cos(\phi - \tau s)) \left(\frac{1 - r}{\delta g(s, \phi)}\right), \; 0, \; \Lambda \tau r \left(\frac{1 - r}{\delta g(s, \phi)}\right) \right) & \quad \text{if} \quad 1 - \delta g(s, \phi) \leq r \leq 1.
\end{cases} \nonumber \\
\label{Lower bound on the volume flow rate: the background flow}
\end{eqnarray}
Here, $\Lambda$ is a constant that will be adjusted later to optimize the bound, 
\begin{eqnarray}
\delta = \frac{1}{\Rey},
\label{Lower bound on the volume flow rate: def of delta that's right}
\end{eqnarray} 
and $g(s, \phi)$ is a non-zero bounded differentiable function of $s$ and $\phi$, that satisfies
\begin{eqnarray}
0 < g_l \leq g(s, \phi) \leq g_u \quad \text{and} \quad \left|\frac{\partial g}{\partial s}\right|, \left|\frac{\partial g}{\partial \phi}\right| \leq g^\prime_u \quad \forall \; s \in [0, s_p], \phi \in [0, 2 \upi],
\label{Lower bound on the volume flow rate: constraint on g}
\end{eqnarray}
where $g_l$, $g_u$, and $g^\prime_u$ are constants independent of $\Rey$. The region $1 - \delta g(s, \phi) \leq r \leq 1$ is the boundary layer  denoted as 
\begin{eqnarray}
 \Omega_\delta = \{(s, r, \phi) | s \in [0, s_p], \phi \in [0, 2 \upi], 1 - \delta g(s, \phi) \leq r \leq 1\},
\end{eqnarray}
and the function $g(s, \phi)$ represents the shape of the boundary layer which will be determined later as part of the analysis to optimize the bound. Physically, (\ref{Lower bound on the volume flow rate: constraint on g}) means that the thickness of the boundary layer is everywhere non-zero and finite, and it varies smoothly. Figure (\ref{The background flow illustration}) shows a color map of the streamwise component $U_s$ of the background flow (\ref{Lower bound on the volume flow rate: the background flow}). It can be easily verified that the background flow field (\ref{Lower bound on the volume flow rate: the background flow}) satisfies the no-slip and impermeable boundary conditions on the pipe surface. Meanwhile, the divergence-free condition on the background flow enforces
\begin{eqnarray}
g(s, \phi) = g(\phi - \tau s),
\label{Lower bound on the volume flow rate: div-free constraint on g}
\end{eqnarray}
which constrains the choice of $g$. See Appendix \ref{s r phi coordinate system} for the calculation of divergence of a vector in the $(s, r, \phi)$ coordinate system. Also, note that for this choice of $\bs{U}$, in  the bulk region ($0 \leq r \leq 1 - \delta g(s, \phi)$), we have
\begin{eqnarray}
\bnabla \bs{U}_{\text{sym}} = \mathsfbi{0} 
\label{Lower bound on the volume flow rate: rigid body condition}
\end{eqnarray}
(see Appendix \ref{Some useful calculations}) the reason being is that in this region $\bs{U}$ is really a rigid body flow as viewed from some inertial frame of reference  (see \S\ref{Discussion and concluding remarks} Discussion and Concluding Remarks).  Although we can obtain a bound on the flow rate with a constant boundary layer thickness, this choice does not provide the optimal bound as a function of the pipe's curvature $\kappa$ and torsion $\tau$. Unlike the case of planar geometries, the choice of the background flow (\ref{Lower bound on the volume flow rate: the background flow}) is not uniform in the bulk region. As can be seen in figure (\ref{The background flow illustration}), the magnitude of the streamwise component $U_s$ of the background flow in the bulk region varies and is higher towards the outer edge O of the pipe than the inner edge. As such, a constant boundary layer thickness is not necessarily the optimal choice for every $\kappa$ and $\tau$. Therefore, it is natural to choose a variable boundary layer thickness (which is more general than the choice of constant boundary layer thickness) since our goal is to optimize bounds simultaneously for different curvature and torsion. Furthermore, in the process of obtaining bounds, we complement this choice of variable boundary layer thickness with inequalities suitably constructed (from standard analysis inequalities) to achieve this goal.
In the forthcoming analysis, we will be interested in obtaining a bound in the limit of high Reynolds number and therefore will be frequently making use of the fact that $\Rey \gg 1$ or $\delta \ll 1$ to retain only the leading order terms.

\subsection{The spectral constraint}
In this subsection, we use analysis techniques to obtain a condition under which the spectral constraint (\ref{The background method formulation: spectral constraint}) is satisfied.   In what follows, we shall make use of a crucial inequality, whose proof is given in Appendix \ref{Proof of two inequalities}. 
\begin{customlemma}{1} Let $w:\Omega_\delta \to \mathbb{R}$ be a square integrable function such that $w(s, 1, \phi) = 0 $ for all $0 \leq s \leq s_p$ and $0 \leq \theta \leq 2 \upi$, then the following statement is true
\begin{eqnarray}
\int_{\Omega_{\delta}} \sigma w^2 \; d \bs{x} \;   \leq  \; \frac{\delta^2}{2}\int_{\Omega_{\delta}} \sigma (s, 1, \phi) g^2(s, \phi) \left(\frac{\partial w}{\partial r}\right)^2 \; d \bs{x} + O(\delta^3) \; ||\bnabla w||_2^2.
\label{Appendix inequalities: Inequality 3}
\end{eqnarray}
Here, $\sigma:\Omega_\delta \to \mathbb{R}$ is a positive bounded $O(1)$ function that satisfies
\begin{eqnarray}
|\sigma (s, r, \phi) - \sigma (s, 1, \phi)| = O(\delta) \quad \text{for } (s, r, \phi) \in \Omega_{\delta}.
\end{eqnarray}
\end{customlemma}
For convenience, we make use of the big O notation $O(\cdot)$. Let $m$ and $n$ be two functions, then in this notation, writing $m(\delta) = O(n(\delta))$ means that there exists two positive constants $C > 0$ and $\delta_0 > 0$ such that $|m(\delta)| \leq C |n(\delta)|$ whenever $0 \leq \delta < \delta_0$.

 We start by obtaining a bound on $I$  as defined in (\ref{The background method formulation: def H(v)}). Making use of (\ref{Lower bound on the volume flow rate: rigid body condition}) leads to
\begin{eqnarray}
I = \int_{\Omega} (\bs{v} \bcdot \bnabla \bs{U}_{\text{sym}}) \bcdot \bs{v} \; d \bs{x} = \int_{\Omega_{\delta}} (\bs{v} \bcdot \bnabla \bs{U}_{\text{sym}}) \bcdot \bs{v} \; d \bs{x}.
\label{Lower bound on the volume flow rate: def of I using Omega delta}
\end{eqnarray}
The following inequality is obtained by substituting $\bnabla \bs{U}_{\text{sym}}$ (use (\ref{Appendix calculations: grad U boundary layer}) from Appendix \ref{Some useful calculations} for the calculation of $\bnabla \bs{U}_{\text{sym}}$) into (\ref{Lower bound on the volume flow rate: def of I using Omega delta})
\begin{eqnarray}
|I| \leq &&  \int_{\Omega_\delta} \xi_1(s, r, \phi) |v_s| |v_r| \; d \bs{x}  + \int_{\Omega_\delta} \xi_2(s, r, \phi) |v_{\phi}| |v_r|\; d \bs{x} \nonumber \\
&& + \int_{\Omega_\delta} \xi_3 v_s^2 d \bs{x}  + \int_{\Omega_\delta} \xi_4  v_\phi^2 \; d \bs{x} + \int_{\Omega_\delta} \xi_5  |v_s| |v_\phi| \; d \bs{x} ,
\label{Lower bound on the volume flow rate: bound I}
\end{eqnarray} 
where
\begin{eqnarray}
\xi_1(s, r, \phi) = \frac{\Lambda (1 - r \kappa \cos(\phi - \tau s)) }{\delta g}, \qquad 
\xi_2(s, r, \phi) = \frac{\Lambda \tau r}{\delta g}, \qquad \qquad \qquad \nonumber \\
\xi_3 = \max_{(s, r, \phi) \in \Omega_\delta}\frac{\Lambda (1-r)}{\delta g^2} \left|\frac{\partial g}{\partial s}\right|, \qquad 
\xi_4 = \max_{(s, r, \phi) \in \Omega_\delta} \frac{\Lambda \tau (1-r)}{\delta g^2} \left|\frac{\partial g}{\partial \phi}\right|, \qquad \qquad \nonumber \\
\xi_5 = \max_{(s, r, \phi) \in \Omega_\delta} \frac{\Lambda (1-r)}{\delta g^2} \left[\frac{\tau r}{(1 - r \kappa \cos(\phi - \tau s))} \left|\frac{\partial g}{\partial s}\right| + \frac{(1 - r \kappa \cos(\phi - \tau s))}{r} \left|\frac{\partial g}{\partial \phi}\right|\right]. \nonumber \\
\end{eqnarray} 
Given that $1-r$ is $O(\delta)$ in the boundary layer, and using the constraints on $g$ and its derivatives from (\ref{Lower bound on the volume flow rate: constraint on g}), implies that $\xi_3$, $\xi_4$, and $\xi_5$ are $O(1)$ constants. Using Young's inequality $|v_s| |v_\phi| \leq (|v_s|^2 + |v_\phi|^2)/2$, the last three integrals in (\ref{Lower bound on the volume flow rate: bound I}) can be replaced by
\begin{eqnarray}
\int_{\Omega_\delta} \left(\xi_3 + \frac{\xi_5}{2}\right) v_s^2 d \bs{x}  + \int_{\Omega_\delta}  \left(\xi_4 + \frac{\xi_5}{2}\right) v_\phi^2 \; d \bs{x}.
\end{eqnarray} 
 An application of  Inequality 1 to these two integrals with $w = v_s$ in the first integral, $w = v_\phi$ in the second integral and taking $\sigma$ to be an $O(1)$ constant in both cases results a bound on $I$ as
\begin{eqnarray}
|I| \leq \int_{\Omega_\delta} \xi_1(s, r, \phi) |v_s| |v_r| \; d \bs{x}  + \int_{\Omega_\delta} \xi_2(s, r, \phi)  |v_{\phi}| |v_r|\; d \bs{x} + O(\delta^2) ||\bnabla \bs{v}||_2^2. \nonumber \\
\label{Lower bound on the volume flow rate: premature bound on I}
\end{eqnarray}
In a similar manner, we obtain bounds on the remaining two integrals in (\ref{Lower bound on the volume flow rate: premature bound on I}). These bounds contribute to the leading order term of the bound on $|I|$; therefore, this time we perform the computation wisely with the intent of optimizing the bound on $|I|$ simultaneously in $\kappa$ and $\tau$. Using the following inequalities (based on Young's inequality) 
in (\ref{Lower bound on the volume flow rate: premature bound on I})
\begin{eqnarray}
|v_s| |v_r| \leq \frac{c_1(s, \phi) |v_s|^2}{2} + \frac{|v_r|^2}{2 c_1(s, \phi)}, \quad |v_{\phi}| |v_r| \leq \frac{c_2(s, \phi) |v_{\phi}|^2}{2} + \frac{|v_r|^2}{2 c_2(s, \phi)},
\label{Lower bound on the volume flow rate: s phi Young's inequality}
\end{eqnarray}
where
\begin{eqnarray}
0 <  c_1(s, \phi)  \quad \text{and} \quad 0 <  c_2(s, \phi), 
\end{eqnarray}
results in
\begin{eqnarray}
|I| \leq \int_{\Omega_\delta} \left[\frac{c_1(s, \phi) \xi_1(s, r, \phi)}{2}\right] |v_s|^2 \; d \bs{x} + \int_{\Omega_\delta} \left[\frac{\xi_1(s, r, \phi)}{2 c_1(s, \phi)} + \frac{\xi_2(s, r, \phi)}{2 c_2(s, \phi)}\right] |v_r|^2 \; d \bs{x} && \nonumber \\
 + \int_{\Omega_\delta} \left[\frac{c_2(s, \phi) \xi_2(s, r, \phi)}{2}\right] |v_\phi|^2 \; d \bs{x} + O(\delta^2) ||\bnabla \bs{v}||_2^2. &&
\label{Lower bound on the volume flow rate: premature intermediate bound on I}
\end{eqnarray}
We apply the Inequality 1 to the three integrals in (\ref{Lower bound on the volume flow rate: premature intermediate bound on I}) with $w=v_s, v_r$, and $v_\phi$ and taking $\sigma$ to be the corresponding terms in the square brackets times $\delta$, which results in
\begin{eqnarray}
|I| && \leq \frac{\Lambda \delta}{4} \left[ \int_{\Omega_{\delta}} p_1 \left(\frac{\partial v_s}{\partial r}\right)^2 \; d \bs{x} + \int_{\Omega_{\delta}} p_2 \left(\frac{\partial v_r}{\partial r}\right)^2 \; d \bs{x} + \int_{\Omega_{\delta}} p_3 \left(\frac{\partial v_{\phi}}{\partial r}\right)^2 \; d \bs{x}\right] + O(\delta^2) ||\bnabla \bs{v}||_2^2, \nonumber \\
\label{Lower bound on the volume flow rate: bound on I in p1 p2 p3 1}
\end{eqnarray}
where
\begin{eqnarray}
&& p_1 =  (1-\kappa \cos(\phi - \tau s))  g(s, \phi) c_1(s, \phi), \nonumber \\ 
&& p_2 = \frac{ (1-\kappa \cos(\phi - \tau s))  g(s, \phi) }{ c_1(s, \phi)} + \frac{ \tau  g(s, \phi) }{ c_2(s, \phi)}, \nonumber \\
&& p_3 =  \tau  g(s, \phi) c_2(s, \phi).
\label{Lower bound on the volume flow rate: p1 p2 p3 1}
\end{eqnarray}
We now choose the functions $g(s, \phi)$, $c_1(s, \phi)$, and $c_2(s, \phi)$ so that $p_1$, $p_2$, and $p_3$ are constants. For this choice, the bound on $I$ can be written as
\begin{eqnarray}
|I| && \leq  \frac{\Lambda \delta}{4} \max \{p_1, p_2, p_3\} \; ||\bnabla \bs{v}||_2^2 + O(\delta^2) ||\bnabla \bs{v}||_2^2.
\label{Lower bound on the volume flow rate: bound on I in p1 p2 p3 2}
\end{eqnarray}
To optimize the bound, we need
\begin{eqnarray}
p_1 = p_2 = p_3,
\label{Lower bound on the volume flow rate: optimality condition}
\end{eqnarray}  as shown in Appendix \ref{Some useful calculations}. Combining this condition with the requirement that $p_1, p_2,$ and $p_3$ should be constants leads to
\begin{eqnarray}
g(s, \phi) = \frac{g_c}{\sqrt{(1-\kappa \cos(\phi - \tau s))^2 + \tau^2}} \qquad \qquad \qquad \qquad \nonumber \\
c_1(s, \phi) = \sqrt{1 + \frac{\tau^2}{(1-\kappa \cos(\phi - \tau s))^2}}, \quad c_2(s, \phi) = \sqrt{1 + \frac{(1-\kappa \cos(\phi - \tau s))^2}{\tau^2}}
\label{Lower bound on the volume flow rate: g c1 c2}
\end{eqnarray}
with $g_c$ being an $O(1)$ positive constant. Note that the function $g(s, \phi)$ satisfies the constraints (\ref{Lower bound on the volume flow rate: constraint on g}) and (\ref{Lower bound on the volume flow rate: div-free constraint on g}),  where the constants $g_l$, $g_u$, $g_u^\prime$ in (\ref{Lower bound on the volume flow rate: constraint on g}) can be chosen as
\begin{eqnarray}
g_l = \frac{g_c}{\sqrt{(1+\kappa)^2 + \tau^2}}, \quad g_u = \frac{g_c}{\sqrt{(1-\kappa)^2 + \tau^2}}, \quad \text{and }  g_u^\prime =  \frac{2 g_c }{\left[(1-\kappa)^2 + \tau^2\right]^{3/2}}.
\end{eqnarray} 
Combining (\ref{Lower bound on the volume flow rate: p1 p2 p3 1}), (\ref{Lower bound on the volume flow rate: bound on I in p1 p2 p3 2}), and (\ref{Lower bound on the volume flow rate: g c1 c2}) gives a bound on $I$ as
\begin{eqnarray}
|I| \leq \left|\int_{\Omega} \bs{v} \bcdot \bnabla \bs{U} \bcdot \bs{v} \; d \bs{x}\right| \leq \left(\frac{\Lambda g_c \delta}{4} + O(\delta^2)\right) ||\bnabla \bs{v}||_2^2.
\end{eqnarray}
Next, we show that the contribution of term $II$, as defined in (\ref{The background method formulation: def H(v)}), is of higher order in $\delta$ compared to term $I$. First note that for any scalar function $\Psi$
\begin{eqnarray}
II = \int_{\Omega} \bs{U} \bcdot \bnabla \bs{U} \bcdot \bs{v} \; d \bs{x} = \int_{\Omega} (\bs{U} \bcdot \bnabla \bs{U} - \bnabla \Psi) \bcdot \bs{v} \; d \bs{x}
\end{eqnarray}
using the incompressibility of $\bs{v}$, together with the fact that $\bs{v}$ satisfies the homogenous boundary conditions. Then, if we choose $\Psi$ such that $\bs{U} \bcdot \bnabla \bs{U} = \bnabla \Psi$ in $\Omega \setminus \Omega_{\delta}$, namely,
\begin{eqnarray}
\Psi(s, r, \phi) = \Lambda^2 \kappa \cos(\phi - \tau s) \left(r - \frac{r^2}{2} \kappa \cos(\phi - \tau s)\right) - \frac{\Lambda^2 \tau^2 r^2}{2},
\end{eqnarray}
then one can readily check that
\begin{eqnarray}
|(\bs{U} \bcdot \bnabla \bs{U} - \bnabla \Psi)| (\bs{x}) = 
\begin{cases}
0 & \quad \text{if } \bs{x} \in \Omega \setminus \Omega_{\delta} \\
O(1) & \quad \text{if } \bs{x} \in \Omega_{\delta}
\end{cases}.
\label{Lower bound on the volume flow rate: U grad U}
\end{eqnarray}
See Appendix \ref{Some useful calculations} for the calculation of $\bnabla \bs{U}$. Using (\ref{Lower bound on the volume flow rate: U grad U}), we can finally obtain a bound on $II$ as
\begin{eqnarray}
|II| && = \left|\int_{\Omega} \bs{U} \bcdot \bnabla \bs{U} \bcdot \bs{v} \; d \bs{x}\right| = \left|\int_{\Omega} (\bs{U} \bcdot \bnabla \bs{U} - \bnabla \Psi) \bcdot \bs{v} \; d \bs{x}\right|\nonumber \\
\implies |II| && \leq O(1) \int_{\Omega_{\delta}} |\bs{v}| \; d \bs{x} \nonumber \\
&& \leq O(1) \int_{\Omega_{\delta}} |\bs{v}|^2 \; d \bs{x} + O(1) \int_{\Omega_{\delta}} 1 \; d \bs{x} \nonumber \\ 
&& \leq O(\delta^{2}) ||\bnabla \bs{v}||_2^2 + s_p O(\delta).
\end{eqnarray}
We have used Young's inequality to obtain the third line and Inequality 1 to obtain the last line.  Finally,we obtain a bound on $\mathcal{H}(\bs{v})$ defined in (\ref{The background method formulation: def H(v)}) using the triangle inequality and the bounds derived on $I$ and $II$ as
\begin{eqnarray}
\mathcal{H}(\bs{v}) \geq \frac{1}{2 \Rey} ||\bnabla \bs{v}||_2^2 - \left(\frac{\Lambda g_c \delta}{4} + O(\delta^2)\right) ||\bnabla \bs{v}||_2^2 -  s_p O(\delta),
\end{eqnarray}
which implies
\begin{eqnarray}
\mathcal{H}(\bs{v}) + \gamma \geq 0
\end{eqnarray}
as long as
\begin{eqnarray}
g_c \leq \frac{2}{\Lambda} + O(\delta) \quad \text{and} \quad \gamma = s_p O(\delta).
\label{Lower bound on the volume flow rate: gc gamma}
\end{eqnarray}

\subsection{Bound on mean quantities}
\begin{figure}
\centering
 \includegraphics[scale = 0.8]{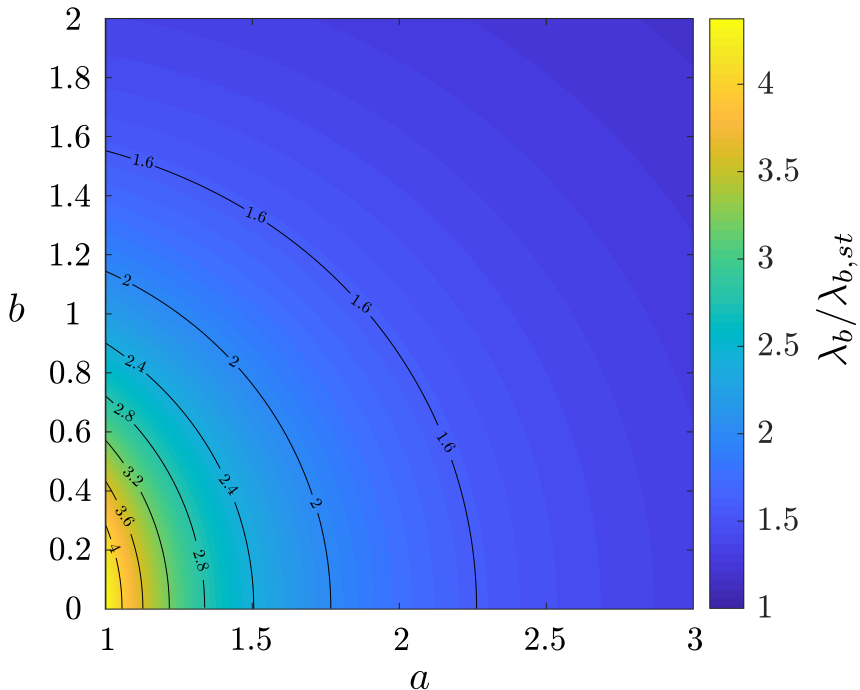}
 \caption{Ratio of the bound on the friction factor for a helical pipe as compared to a straight pipe ($\lambda_b/\lambda_{b, st}$), as a function of the non-dimensional geometric parameters $a$ and $b$ as defined earlier.}
 \label{f/fb plot a b}
\end{figure}

We are now ready to compute the bound on the flow rate. We begin by evaluating the first term on the right-hand-side of (\ref{The background method formulation: bound on Q}) as
\begin{eqnarray}
\int_{\phi = 0}^{2 \upi} \int_{r = 0}^{1} U_s r dr d\phi && = \int_{\phi = 0}^{2 \upi} \int_{r = 0}^{1 - \delta g(s, \phi)} \Lambda (1 - r \kappa \cos(\phi - \tau s)) r dr d\phi \nonumber \\
&& + \int_{\phi = 0}^{2 \upi} \int_{1 - \delta g(s, \phi)}^{1} \Lambda (1 - r \kappa \cos(\phi - \tau s)) \frac{(1-r)}{\delta g} r dr d\phi \nonumber \\
&& = \int_{\phi = 0}^{2 \upi} \int_{r = 0}^{1} \Lambda (1 - r \kappa \cos(\phi - \tau s)) r dr d\phi + O(\delta) \nonumber \\
&& = \upi \Lambda + O(\delta).
\label{Bound on mean quantities: Q RHS 1st}
\end{eqnarray}
Similarly, the second term on the right-hand-side of (\ref{The background method formulation: bound on Q}) is as follows
\begin{eqnarray}
||\bnabla \bs{U}||_2^2 && = \int_{\Omega_\delta} |\bnabla \bs{U}|^2 \;  d \bs{x} + \int_{\Omega \setminus \Omega_\delta} |\bnabla \bs{U}|^2 \;  d \bs{x} \nonumber \\
&& = \int_{s = 0}^{s_p} \int_{\phi = 0}^{2 \upi} \int_{1 - \delta g(s, \phi)}^{1} |\bnabla \bs{U}|^2 \;  h_s h_r h_\phi dr d \phi ds + s_p O(1) \nonumber \\
&& = \frac{2 \upi \Lambda^2 s_p}{\delta \; g_c} I(\kappa, \tau) + s_p O(1),
\label{Bound on mean quantities: Q RHS 2nd}
\end{eqnarray} 
where
\begin{eqnarray}
I(\kappa, \tau) =  \frac{1}{2 \upi}\int_{0}^{2 \upi} \left((1-\kappa \cos \alpha)^2 + \tau^2\right)^{3/2}(1-\kappa \cos \alpha) \; d \alpha.
\label{Bound on mean quantities: def I}
\end{eqnarray}
The integrand in the second line of (\ref{Bound on mean quantities: Q RHS 2nd}) is explicitly calculated in (\ref{Appendix calculations: grad U2}), see Appendix \ref{Some useful calculations}. Using (\ref{Bound on mean quantities: Q RHS 1st}) and (\ref{Bound on mean quantities: Q RHS 2nd}) in (\ref{The background method formulation: bound on Q}), we obtain a bound on the flow rate $Q$ as
\begin{eqnarray}
Q \geq 2 \upi \Lambda - \frac{2 \upi \Lambda^2}{g_c} I(\kappa, \tau) + O(\delta).
\end{eqnarray}
We see that choosing a small value of $g_c$ will make the bound on $Q$ worse. Therefore, to obtain the best possible bound, we choose the largest possible value of $g_c$ which satisfies the constraint (\ref{Lower bound on the volume flow rate: gc gamma}), i.e., 
\begin{eqnarray}
g_c = \frac{2}{\Lambda} + O(\delta).
\end{eqnarray} 
The bound on $Q$ then reads
\begin{eqnarray}
Q \geq 2 \upi \Lambda - \upi \Lambda^3 I(\kappa, \tau) + O(\delta).
\label{Lower bound on the volume flow rate: lower bound on Q almost there}
\end{eqnarray} 
All that remains is to choose $\Lambda$ to maximize this lower bound. The optimal value of $\Lambda$ is given by
\begin{eqnarray}
\Lambda = \sqrt{\frac{2}{3 I(\kappa, \tau)}}.
\label{Lower bound on the volume flow rate: optimal value of lambda half peace}
\end{eqnarray}
Substituting (\ref{Lower bound on the volume flow rate: optimal value of lambda half peace}) into (\ref{Lower bound on the volume flow rate: lower bound on Q almost there}) and using $\delta = 1/\Rey$ as defined earlier in (\ref{Lower bound on the volume flow rate: def of delta that's right}), gives a bound on the flow rate as
\begin{eqnarray}
Q \geq \sqrt{\frac{32 \upi^2}{27 I(\kappa, \tau)}}  +  O(\Rey^{-1}).
\label{Lower bound on the volume flow rate: lower bound on Q}
\end{eqnarray}
\begin{table}
  \begin{center}
\def~{\hphantom{0}}
  \begin{tabular}{lccc}
      Pipe  & Flow rate $Q$ (lower bound)   &  Friction factor $\lambda$ (upper bound) \\[3pt]
       Straight   & $\mathscr{C}$ & $\mathscr{D}$ \\
       Helical   & $\mathscr{C} / \sqrt{I(\kappa, \tau)}$ & $\mathscr{D} I(\kappa, \tau)$ \\
       Toroidal  & $\mathscr{C} \left(1 + 3 \kappa^2 + \frac{3}{8} \kappa^4\right)^{-\frac{1}{2}}$ & $\mathscr{D} \left(1 + 3 \kappa^2 + \frac{3}{8} \kappa^4\right)$ \\
       Helical ($\tau \ll 1$)   & $\mathscr{C} \left(1 + 3 \kappa^2 + \frac{3}{8} \kappa^4\right)^{-\frac{1}{2}} \left(1 - \frac{3 (2 + \kappa^2) \tau^2}{8 + 24 \kappa^2 + 3\kappa^4} \right)$ & $\mathscr{D} \left(1 + 3 \kappa^2 + \frac{3}{8} \kappa^4\right) \left(1 + \frac{6 (2 + \kappa^2) \tau^2}{8 + 24 \kappa^2 + 3 \kappa^4}\right)$ \\
  \end{tabular}
  \caption{The first column shows the pipe type, the second column shows the lower bound on the flow rate $Q$ (\ref{Lower bound on the volume flow rate: lower bound on Q}) and the third column shows the upper bound on the friction factor $\lambda$ (\ref{Bound on mean quantities: upper bound on f}) in different limits of curvature $\kappa$ and torsion $\tau$. In the table $\mathscr{C} = (32 \upi^2 / 27)^{\frac{1}{2}}$ and $\mathscr{D} = 27/8$.}
  \label{Table 1: bounds in different limits}
  \end{center}
\end{table}
Using this lower bound, together with the definition (\ref{The background method formulation: fric frac}) of the friction factor $\lambda$ in terms of $Q$, we finally obtain an upper bound on the friction factor as
\begin{eqnarray}
\lambda \leq \lambda_b = \frac{27}{8} I(\kappa, \tau) +  O(\Rey^{-1}).
\label{Bound on mean quantities: upper bound on f}
\end{eqnarray}
\begin{figure}
\centering
 \includegraphics[scale = 1.0]{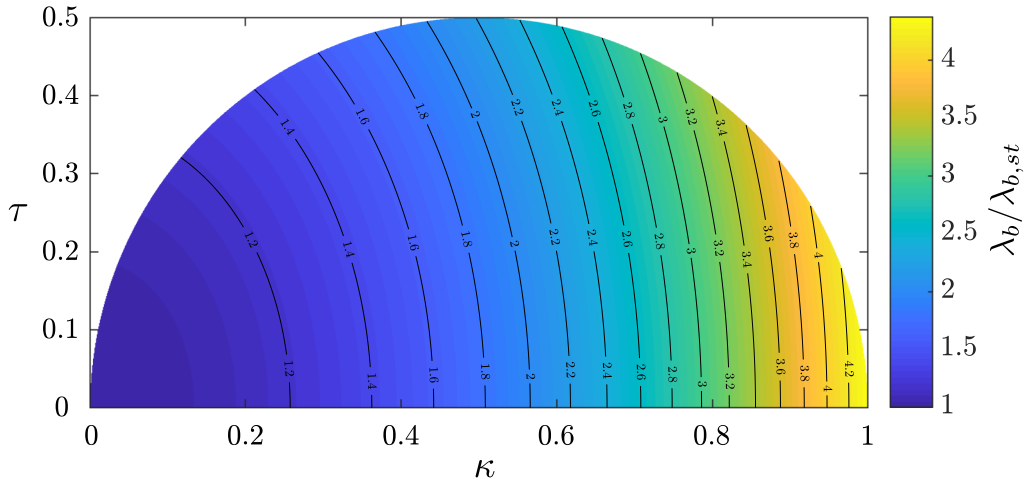}
 \caption{Ratio of the bound on the friction factor for a helical pipe as compared to a straight pipe ($\lambda_b/\lambda_{b, st}$), as a function of curvature $\kappa$ and torsion $\tau$. Here, $\lambda_b$ is given by (\ref{Bound on mean quantities: upper bound on f}) and $\lambda_{b, st} = 27/8$.}
 \label{f/fb kappa tau plot}
\end{figure}
These bounds on the flow rate (\ref{Lower bound on the volume flow rate: lower bound on Q}) and friction factor (\ref{Bound on mean quantities: upper bound on f}) are also valid for a toroidal ($\kappa \neq 0, \tau = 0$)  or a straight ($\kappa = 0, \tau = 0$) pipe. In general, the integral $I(\kappa, \tau)$ cannot be obtained analytically except when $\tau$ is zero or small. Table \ref{Table 1: bounds in different limits} summarizes the bounds derived on the flow rate and friction factor in different limits of curvature and torsion. For a toroidal pipe,  a small radius of curvature has a second-order effect on the bounds (see table \ref{Table 1: bounds in different limits}),  which is also the case  in the exact solution for the flow at low Reynolds number \citep[see][]{dean1928lxxii}. Similarly, for a helical pipe, the effect of  a small torsion is of second-order on our bounds, and as before, this is the case for the steady-state solution at low Reynolds number \citep[see][]{tuttle1990laminar}. More generally, the effect of increasing curvature and torsion is always to decrease the lower bound on the volume flow rate and to increase the  upper bound on the friction factor. For a straight pipe, the bound on the friction factor reduces to $\lambda_{b, st} = 27/8$ which is $12.5$ times larger than the bound  $0.27$ \citep{plasting2005friction}  obtained by solving the variational problem numerically. Figures \ref{f/fb kappa tau plot} and \ref{f/fb plot a b} show a color map of the ratio of the bound on the friction factor for a helical pipe $\lambda_b$ to the bound on the friction factor $\lambda_{b, st}$ for a straight pipe, as a function of $\kappa$, $\tau$ (figure \ref{f/fb kappa tau plot}) and $a$, $b$ (figure \ref{f/fb plot a b}). To avoid a self-intersecting geometry, we restrict to $1 < a < \infty$ and $0 \leq b < \infty $, which corresponds to a semicircular region in $\kappa, \tau$ space (see figure \ref{f/fb kappa tau plot}). The maximum increase in the bound on the friction factor is when the pipe approaches a horn torus ($\kappa = 1, \tau = 0$), which is a factor $35/8 = 4.375$ larger than for the straight pipe. From figure \ref{f/fb plot a b}, we see that with the increase of the non-dimensional helix radius $a$ or pitch $2 \upi b$, the bound on the friction factor approaches that of a straight pipe, as expected.

\section{Discussion and Concluding Remarks}
\label{Discussion and concluding remarks}
In this paper, we used the background method to obtain bounds on the flow rate and friction factor in helical pipe flows. The bounds that we obtained are also valid for toroidal and straight pipes as limiting cases. By choosing a boundary layer whose thickness varies along the circumference of the pipe, we were able to obtain these bounds as a function of pipe geometry.  In particular, we found that the bound on the friction factor varies with curvature $\kappa$ and torsion $\tau$ according to the integral $I(\kappa, \tau)$, defined in (\ref{Bound on mean quantities: def I}), whose value is one for the straight pipe, i.e. $I(0, 0) = 1$. 

\begin{figure}
\centering
\begin{tabular}{lc}
\begin{subfigure}{0.5\textwidth}
\centering
 \includegraphics[scale = 0.9]{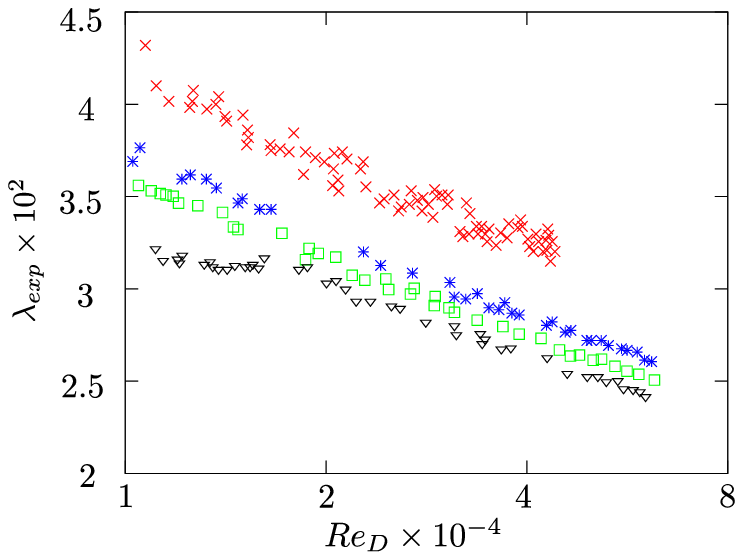}
\caption{}
\end{subfigure} &
\begin{subfigure}{0.5\textwidth}
\centering
 \includegraphics[scale = 0.9]{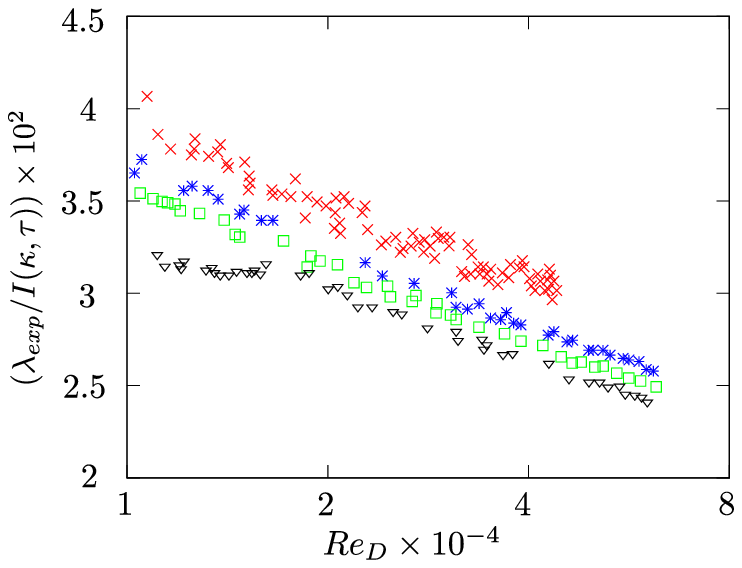}
\caption{}
\end{subfigure}
\end{tabular}
 \caption{(a) Data from \citet{cioncolini2006experimental}, showing the friction factor ($\lambda_{exp}$) as a function of Reynolds number for four different helical pipes: (i) $\kappa = 0.028, \; \tau = 0.49\times 10^{-3}$ ($\color{black} {\triangledown}$), (ii) $\kappa = 0.042, \; \tau = 1.87\times 10^{-3}$($\color{green} {\Box}$), (iii)  $\kappa = 0.059, \; \tau = 2.97\times 10^{-3}$ ($\color{blue} {\ast}$), and (iv) $\kappa = 0.143, \; \tau = 11.4\times 10^{-3}$ ($\color{red} \boldsymbol{\times}$). (b) Scaled friction factor ($\lambda_{exp}/I(\kappa, \tau)$) as a function of Reynolds number for the same four helical pipes.}
 \label{Comparison experiment}
\end{figure}

The bound that we obtained on the friction factor is independent of the Reynolds number. However, it is a known property of wall-bounded flows in smooth geometries that the friction factor decreases as $(\log \Rey)^2$, as predicted using the logarithmic friction law \citep{tennekes1972first}. Therefore, it appears that our bound, in terms of Reynolds number scaling is off by factor of $(\log \Rey)^2$, a situation similar to previous applications of the background method to shear and pressure-driven flows \citep{PhysRevLett.69.1648, PhysRevE.49.4087, PhysRevE.51.3192, plasting2003improved}.  Despite being unable to capture the correct scaling in terms of Reynolds number, one may ask whether the geometrical scaling $I(\kappa, \tau)$ in the bound on the friction factor (\ref{Bound on mean quantities: upper bound on f}) correctly captures the variation of $\lambda$ on the pipe geometry, as observed in the experiments. Assuming that is the case, then one would expect that the experimental data for the friction factor ($\lambda_{exp}$), when scaled with the integral $I(\kappa, \tau)$, should only be a function of the Reynolds number. Here, we test this hypothesis on data from carefully conducted experiments by \citet{cioncolini2006experimental} for flows in helical pipes with negligible torsion. The results are shown in figure \ref{Comparison experiment}. In reporting these results, we have used the Reynolds number based on the pipe diameter ($\Rey_D = 2 \Rey$), to be consistent with the literature. From figure \ref{Comparison experiment}, we see that our scaling, being second-order in the curvature $\kappa$, has a negligible effect on the pipes with small curvature ratios $\kappa = 0.028, 0.042, 0.059$.   As such, the rescaled data $\lambda_{exp} / I(\kappa, \tau)$ looks almost identical to the original data, and does not collapse on a universal curve, contrary to our expectation.  It thus appears that, for the range of Reynolds number considered in figure \ref{Comparison experiment}, the curvature has a first-order effect on the friction factor in the experiments,  as opposed to the second-order effect predicted by our bound (see table \ref{Table 1: bounds in different limits}). On a more positive note, the effect becomes qualitatively noticeable for $\kappa = 0.143$, and the rescaled data is more compact than the original data,  suggesting that the dependence on the curvature $\kappa$ given by the scaling $I(\kappa, \tau)$ at least has the right sign. 

There could be several possible reasons for the discrepancy between the data and the theoretical bound. First, there is further improvement possible in our analysis to capture the geometrical dependence of the friction factor better. One way to find that out would be to compute numerically the optimal bound on, e.g. the  friction factor, similar to other studies of the background method \citep{plasting2003improved, wen2013computational, wen2015time, fantuzzi2015construction, fantuzzi2016optimal, fantuzzi2018boundsB, tilgner2017bounds, tilgner2019time} and then see if this optimal bound better accounts for the experimental data. The second possible reason for the discrepancy is that the Reynolds numbers achieved in the experiments are not high enough for our scalings to apply yet. Indeed, the critical Reynolds number for instability for a torus with curvature ratio $\kappa \approx 0.1$ is $Re_{D,c} \approx 3500$ \citep{canton2016modal}, which is higher than that of a straight pipe $Re_{D,c} \approx 2040$ \citep{avila2011onset}.  Also, the transition for $\kappa \gtrapprox 0.028$ is supercritical \citep{kuhnen2015subcritical}, suggesting that the flow structure slowly becomes more complex with increasing Reynolds number, only becoming fully developed turbulence at much higher Reynolds number. As a result, we believe that the maximum Reynolds number achieved by \citet{cioncolini2006experimental} ($\Rey_D \approx 6 \times 10^4$), may not be high enough for our scalings to apply. 

\begin{figure}
\centering
\begin{tabular}{lc}
\begin{subfigure}{0.5\textwidth}
\centering
 \includegraphics[scale = 1.2]{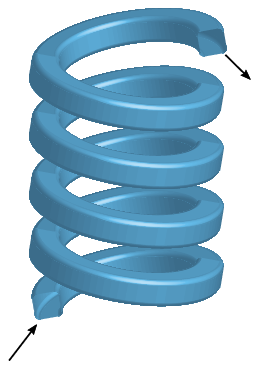}
\caption{}
\end{subfigure} &
\begin{subfigure}{0.5\textwidth}
\centering
 \includegraphics[scale = 2.0]{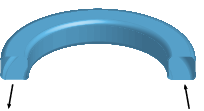}
\caption{}
\end{subfigure}\vspace{1cm}\\
\begin{subfigure}{0.5\textwidth}
\centering
 \includegraphics[scale = 0.30]{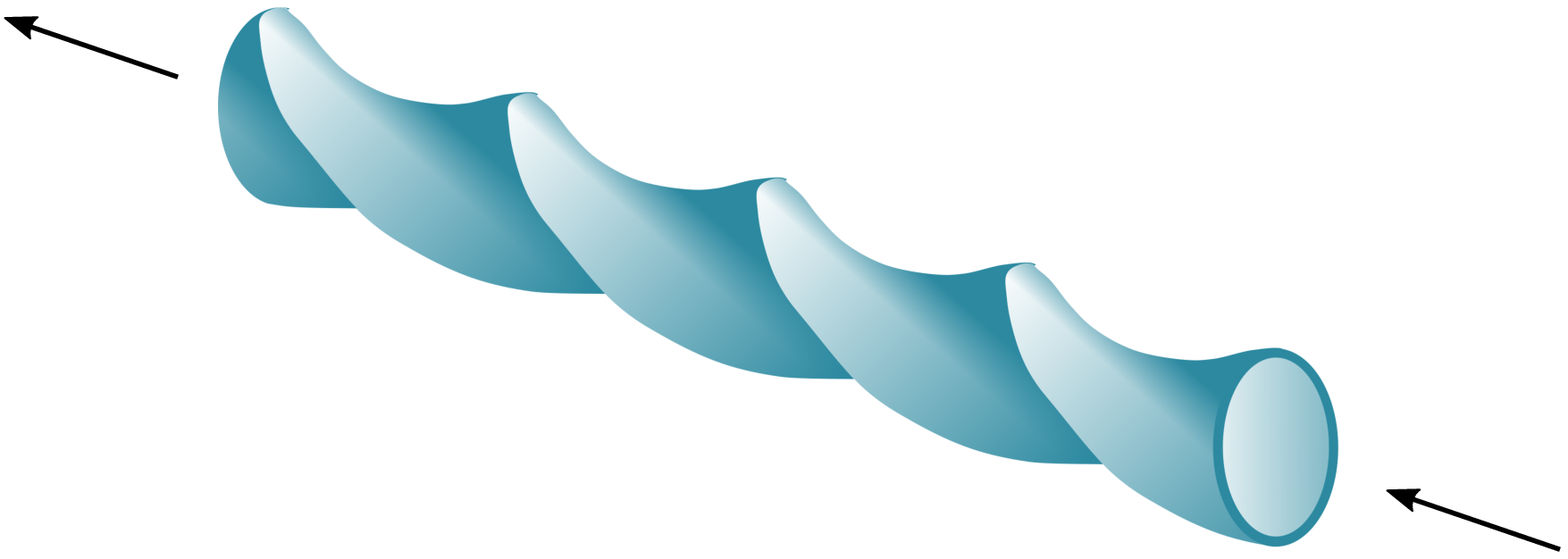}
\caption{}
\end{subfigure} &
\begin{subfigure}{0.5\textwidth}
\centering
 \includegraphics[scale = 1.0]{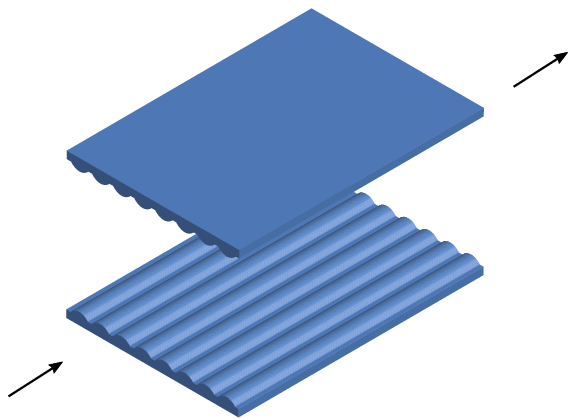}
\caption{}
\end{subfigure}\vspace{1cm}\\
\begin{subfigure}{0.5\textwidth}
\centering
 \includegraphics[scale = 1.0]{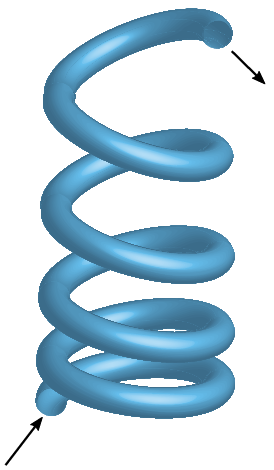}
\caption{}
\end{subfigure} &
\begin{subfigure}{0.5\textwidth}
\centering
 \includegraphics[scale = 2.0]{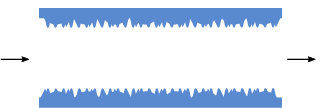}
\caption{}
\end{subfigure}
\end{tabular}
 \caption{Pressure driven flow (a) through a helical pipe with a square cross-section, (b) through a toroidal pipe with a square cross-section, (c) through an axially twisted pipe with an elliptical cross-section, (d) between grooved walls where the grooves are aligned in the direction of the pressure gradient, (e) through a helical pipe with varying pitch, and (f) between rough walls (two-dimensional view). Arrows indicate the direction of the mean flow.}
 \label{examples examples examples examples}
\end{figure}

Beyond the fact that the bounds and the data do not agree as well as we could have hoped for, the technique used in this paper to derive formal bounds for such complex geometry deserves to be discussed. The feasibility of the background method relies on the existence of a divergence-free background flow field $\bs{U}$, which satisfies the same boundary conditions as
the full flow $\bs{u}$ and for which $\mathcal{H}(\bs{v}) + \gamma$ is positive semi-definite, i.e. the spectral constraint (\ref{The background method formulation: spectral constraint}) is satisfied. The situation becomes particularly difficult at high Reynolds number when the only undoubtedly positive term in $\mathcal{H}(\bs{v})$ (see term $III$ in equation \ref{The background method formulation: def H(v)}) becomes small. However, for an $O(1)$ background flow $\bs{U}$ for which $\bnabla \bs{U}_{\text{sym}}$ is zero in bulk and is  of $O(\delta^{-1})$ in a $O(\delta)$ thick boundary layer near the surface, it is possible to show that $\mathcal{H}(\bs{v}) + \gamma$ is positive semi-definite, as done in the present study and in several other studies of the background method \citep{PhysRevE.49.4087, marchioro1994remark, PhysRevE.51.3192, wang1997time}. One may therefore generally ask under which circumstances can such a background flow $\bs{U}$ exist. We start by making two observations. First, that an $O(1)$ change in the velocity field in a boundary layer of thickness $\delta$ leads to $|\bnabla \bs{U}_{\text{sym}}| = O(\delta^{-1})$ in the boundary layer. Second, that a divergence-free flow field $\bs{V}$ for which $\bnabla \bs{V}_{\text{sym}} = \mathsfbi{0}$ is given by
\begin{eqnarray}
\bs{V}(\bs{x}) = \mathsfbi{A} \bs{x} + \bs{V}_0,
\label{Discussion: possible BF}
\end{eqnarray}
where $\mathsfbi{A}$ is a constant skew-symmetric tensor and $\bs{V}_0$ is a constant velocity field. The flows given by (\ref{Discussion: possible BF}) include uniform flow, and rigid body flow. These two observations tell us that for a problem  with prescribed tangential-velocity boundary conditions at  impermeable boundaries where the boundaries have the shape of streamtubes of the flow field given by (\ref{Discussion: possible BF}), it is  always possible to find a background flow $\bs{U}$ for which $\bnabla \bs{U}_{\text{sym}} = \mathsfbi{0}$ in  the bulk and is $O(\delta^{-1})$ in a $O(\delta)$ thick boundary layer near the surface. Indeed, this is done by choosing $\bs{U} = \bs{V}$ in the bulk and by adjusting the tangential component of $\bs{U}$ by an $O(1)$ in the $O(\delta)$ boundary layer to satisfy the prescribed tangential-velocity boundary conditions. Note that these arguments apply to both pressure-driven flow and surface-velocity-driven flow problems, as they both involve the same spectral constraint (\ref{The background method formulation: def H(v)}). \citet{wang1997time} considered the general problem of bounding the energy dissipation for a flow driven by an imposed tangential velocity at the boundaries in an arbitrary bounded domain when $\mathsfbi{A} = \mathsfbi{0}$ and $\bs{V}_0 = \bs{0}$. Considering the more general cases where $\mathsfbi{A}$ and $\bs{V}_0$ are non-zero enables us to extend the class of problems, as demonstrated in this paper. 

Figure \ref{examples examples examples examples} shows some examples of pressure-driven flows where the criterion mentioned in the last paragraph can or cannot be applied. Although we considered the cross-section of the pipe to be circular in this paper, in general, a bound can be obtained on the friction factor for a helical, toroidal, or a straight pipe of any cross-section. Figure \ref{examples examples examples examples}a and figure \ref{examples examples examples examples}b, for example, show a helical pipe and a toroidal pipe with a square cross-section. These two cases fall under the case when $\mathsfbi{A} \neq \mathsfbi{0}, \bs{V}_0 \neq \bs{0}$ and  $\mathsfbi{A} \neq \mathsfbi{0}, \bs{V}_0 = \bs{0}$, respectively. Figure \ref{examples examples examples examples}c shows a pressure-driven flow through an axially twisted pipe with an elliptical cross-section. According to the criterion mentioned in the last paragraph, the background method can be applied  to this example with $\mathsfbi{A} \neq \mathsfbi{0}, \bs{V}_0 \neq \bs{0}$. Further, we can use the background method with  $\mathsfbi{A} = \mathsfbi{0}, \bs{V}_0 \neq \bs{0}$ in case of pressure-driven flow between grooved walls (figure \ref{examples examples examples examples}d) as long as the grooves are aligned in the direction of the pressure gradient. However, for pressure-driven flow through a helical pipe with varying pitch (figure \ref{examples examples examples examples}e) or pressure-driven flow between rough walls (figure \ref{examples examples examples examples}f), there is no choice of $\mathsfbi{A}$ and  $\bs{V}_0$ which works.

The criterion we have mentioned is, so far, a sufficient criterion for the applicability of the background method. Whether this criterion is also a necessary one remains to be determined. The answer to that question is fundamental since it would provide definite guidance about which problems can and cannot be tackled using the background method.

\section*{Acknowledgement}
A.K. acknowledges the support from the Dean's fellowship and Regents' fellowship from the Baskin School of Engineering at UC Santa Cruz, from the GFD fellowship program 2019 at the Woods Hole Oceanographic Institution, and from NSF AST 1814327. Many thanks to P. Garaud and anonymous referees for providing useful comments, which helped to improve the quality of the paper.

\section*{Declaration of interests}
The author reports no conflict of interest.

\appendix
\section{The $(s, r, \phi)$ coordinate system}
\label{s r phi coordinate system}
An infinitesimal displacement $d\bs{x}$ in the $(s, r, \phi)$ coordinate system can be written as
\begin{eqnarray}
d \bs{x} = h_s ds \; \bs{e}_s + h_r dr \; \bs{e}_r + h_{\phi} d\phi \; \bs{e}_\phi,
\end{eqnarray}
where the scale factors are
\begin{eqnarray}
h_s = (1 - r \kappa \cos(\phi - \tau s)), \quad h_r = 1, \quad h_{\phi} = r.
\end{eqnarray}
The gradient of a scalar field $\Psi$ in the $(s, r, \phi)$ coordinate system is given by
\begin{eqnarray}
\bnabla \Psi = \frac{1}{h_s} \frac{\partial \Psi}{\partial s} \bs{e}_s + \frac{1}{h_r} \frac{\partial \Psi}{\partial r} \bs{e}_r + \frac{1}{h_\phi} \frac{\partial \Psi}{\partial \phi} \bs{e}_\phi.
\end{eqnarray}
The divergence of a vector field $\bs{q} = (q_s, q_r, q_\phi)$ in the $(s, r, \phi)$ coordinate system is 
\begin{eqnarray}
\bnabla \bcdot \bs{q} = \frac{1}{h_s h_r h_{\phi}}\left[\frac{\partial h_r h_\phi q_s}{\partial s} + \frac{\partial h_\phi h_s q_r}{\partial r} + \frac{\partial h_s h_r q_\phi}{\partial \phi}\right].
\end{eqnarray}
Finally, the gradient of a vector  $\bs{q} = (q_s, q_r, q_\phi)$ in the $(s, r, \phi)$ coordinate system is written as
\begin{eqnarray}
\bnabla \bs{q} =  && \left(\frac{1}{h_{s}} \frac{\partial q_{s}}{\partial s} + \frac{q_{r}}{h_{s} h_{r}} \frac{\partial h_{s}}{\partial r} + \frac{q_{\phi}}{h_{s} h_{\phi}} \frac{\partial h_{s}}{\partial \phi} \right) \bs{e}_s \bs{e}_s + \left(\frac{1}{h_{s}} \frac{\partial q_{r}}{\partial s} - \frac{q_{s}}{h_{s} h_{r}} \frac{\partial h_{s}}{\partial r}\right) \bs{e}_s \bs{e}_r \nonumber \\ && + \left(\frac{1}{h_{s}} \frac{\partial q_{\phi}}{\partial s} - \frac{q_{s}}{h_{s} h_{\phi}} \frac{\partial h_{s}}{\partial \phi}\right) \bs{e}_s \bs{e}_\phi +
\left(\frac{1}{h_{r}} \frac{\partial q_{s}}{\partial r} - \frac{q_{r}}{h_{r} h_{s}} \frac{\partial h_{r}}{\partial s}\right) \bs{e}_r \bs{e}_s \nonumber \\ &&  + \left(\frac{1}{h_{r}} \frac{\partial q_{r}}{\partial r} + \frac{q_{s}}{h_{r} h_{s}} \frac{\partial h_{r}}{\partial s} + \frac{q_{\phi}}{h_{r} h_{\phi}} \frac{\partial h_{r}}{\partial \phi}\right) \bs{e}_r \bs{e}_r +
  \left(\frac{1}{h_{r}} \frac{\partial q_{\phi}}{\partial r} - \frac{q_{r}}{h_{r} h_{\phi}} \frac{\partial h_{r}}{\partial \phi}\right) \bs{e}_r \bs{e}_\phi \nonumber \\ &&
+ \left(\frac{1}{h_{\phi}} \frac{\partial q_{s}}{\partial \phi} - \frac{q_{\phi}}{h_{\phi} h_{s}} \frac{\partial h_{\phi}}{\partial s}\right) \bs{e}_\phi \bs{e}_s + 
\left(\frac{1}{h_{\phi}} \frac{\partial q_{r}}{\partial \phi} - \frac{q_{\phi}}{h_{\phi} h_{r}} \frac{\partial h_{\phi}}{\partial r}\right) \bs{e}_\phi \bs{e}_r
  \nonumber \\ && + \left(\frac{1}{h_{\phi}} \frac{\partial q_{\phi}}{\partial \phi} + \frac{q_{r}}{h_{\phi} h_{r}} \frac{\partial h_{\phi}}{\partial r} + \frac{q_{s}}{h_{\phi} h_{s}} \frac{\partial h_{\phi}}{\partial s}\right) \bs{e}_\phi \bs{e}_\phi .
\end{eqnarray}

\section{A few useful inequalities}
\label{Proof of two inequalities}
\begin{customlemma}{0}
Let $w:[1 - \delta g(s, \phi), 1] \to \mathbb{R}$ be a square integrable function such that $w(1) = 0 $, then the following inequality holds
\begin{eqnarray}
w^2(r) \leq \left(\frac{1-r}{1-\kappa \cos(\phi - \tau s)} + O(\delta^2)\right) \int_{1 - \delta g(s, \phi)}^{1} \left(\frac{\partial w}{\partial r^\prime}\right)^2 (1 - r^\prime \kappa \cos(\phi - \tau s)) r^\prime dr^\prime \nonumber \\
\label{Appendix inequalities: Inequality 0 better}
\end{eqnarray}
for given $s$, and $\phi$. Here, $r \in [1 - \delta g(s, \phi), 1]$.
\end{customlemma}
\proof
For $r \in [1 - \delta g(s, \phi), 1]$, using the fundamental theorem of calculus and the Cauchy--Schwarz inequality, the following inequality holds
\begin{eqnarray}
&& w^2(r) = \left|\int_{1}^{r} \frac{d w}{d r^\prime} dr^\prime \right|^2 \nonumber \\
&& \leq \left(\int_{r}^{1} \frac{1}{(1 - r^\prime \kappa \cos(\phi - \tau s)) r^\prime} dr^\prime\right) \left(\int_{1 - \delta g(s, \phi)}^{1} \left(\frac{d w}{d r^\prime}\right)^2 (1 - r^\prime \kappa \cos(\phi - \tau s)) r^\prime dr^\prime \right). \nonumber \\
\end{eqnarray}
As mentioned earlier, the curvature satisfies $\kappa < 1$ and since $r^\prime \leq 1$ in the above expression, we have $(1 - r^\prime \kappa \cos(\phi - \tau s)) < 1$. Therefore, the integrands in both integrals are positive. Finally, using the fact that
$$ \left| \frac{1}{(1 - r^\prime \kappa \cos(\phi - \tau s)) r^\prime}  - \frac{1}{1 - \kappa \cos(\phi - \tau s)} \right| = O(\delta) $$
when $r^\prime \in [1 - \delta g(s, \phi), 1]$ completes the proof.
\QEDB
\\

\begin{customlemma}{1}
Let $w:\Omega_\delta \to \mathbb{R}$ be a square integrable function such that $w(s, 1, \phi) = 0 $ for all $0 \leq s \leq s_p$ and $0 \leq \theta \leq 2 \upi$, then the following statement is true
\begin{eqnarray}
\int_{\Omega_{\delta}} \sigma w^2 \; d \bs{x} \;   \leq  \; \frac{\delta^2}{2}\int_{\Omega_{\delta}} \sigma (s, 1, \phi) g^2(s, \phi) \left(\frac{\partial w}{\partial r}\right)^2 \; d \bs{x} + O(\delta^3) \; ||\bnabla w||_2^2.
\label{Appendix inequalities: Inequality 3}
\end{eqnarray}
Here, $\sigma:\Omega_\delta \to \mathbb{R}$ is a positive bounded $O(1)$ function that satisfies
\begin{eqnarray}
|\sigma (s, r, \phi) - \sigma (s, 1, \phi)| = O(\delta) \quad \text{for } (s, r, \phi) \in \Omega_{\delta}.
\end{eqnarray}
\end{customlemma}
\proof
The calculation is as follows:
\begin{eqnarray}
\int_{\Omega_{\delta}} \sigma w^2 \; d \bs{x} && = \int_{s = 0}^{s_p} \int_{\phi = 0}^{2 \upi} \int_{r = 1 - \delta g(s, \phi)}^{1} \sigma w^2 \; h_s h_r h_{\phi} \; dr d\phi ds \nonumber \\
&& \leq \int_{s = 0}^{s_p} \int_{\phi = 0}^{2 \upi} \int_{r = 1 - \delta g(s, \phi)}^{1} \sigma \left[ \left(\frac{1-r}{1-\kappa \cos(\phi - \tau s)} + O(\delta^2)\right) \right. \nonumber \\
&& \left. \times \int_{1 - \delta g(s, \phi)}^{1} \left(\frac{\partial w}{\partial r^\prime}\right)^2 (1 - r^\prime \kappa \cos(\phi - \tau s)) r^\prime dr^\prime \right]  h_s h_r h_{\phi} \; dr d\phi ds.
\label{Appendix inequalities: Omega del w^2}
\end{eqnarray}
Note that Inequality 0 was used in the second line.  For $(s, r, \phi) \in \Omega_\delta$, with an application of the triangle inequality we have
\begin{eqnarray}
&& |\sigma h_s h_r h_\phi - (1-\kappa \cos(\phi - \tau s))\sigma (s, 1, \phi)| \nonumber \\
&& \leq |\sigma (s, r, \phi) - \sigma (s, 1, \phi)|  \max_{(s, r, \phi) \in \Omega_{\delta}}(h_s h_r h_\phi) + |h_s h_r h_\phi  - (1-\kappa \cos(\phi - \tau s))| \sigma (s, 1, \phi). \nonumber \\
\end{eqnarray}
Noting that $$|\sigma (s, r, \phi) - \sigma (s, 1, \phi)| = O(\delta) \quad \text{and} \quad |h_s h_r h_\phi  - (1-\kappa \cos(\phi - \tau s))| = O(\delta), $$
when $(s, r, \phi) \in \Omega_{\delta}$, and 
$$ \max_{(s, r, \phi) \in \Omega_{\delta}}(h_s h_r h_\phi) = O(1) \quad \text{and} \quad \sigma (s, 1, \phi) = O(1), \quad $$ leads to
\begin{eqnarray}
|\sigma h_s h_r h_\phi - (1-\kappa \cos(\phi - \tau s))\sigma (s, 1, \phi)| = O(\delta) \quad \text{for } (s, r, \phi) \in \Omega_{\delta}.
\label{Appendix inequalities: a result of peace}
\end{eqnarray}
Using (\ref{Appendix inequalities: a result of peace}) in (\ref{Appendix inequalities: Omega del w^2}) and performing the integration in $r$ leads to the desired result.
\QEDB

\section{Some useful calculations}
\label{Some useful calculations}
\subsection{Calculation of $\bnabla \bs{U}$}
\label{Some useful calculations: grad U}
For the background flow given by (\ref{Lower bound on the volume flow rate: the background flow}), for $\bs{x} \in \Omega \setminus \Omega_\delta$, we have
\begin{eqnarray}
\bnabla \bs{U} = 
\Lambda \kappa \cos(\phi - \tau s) \bs{e}_s \bs{e}_r  -\Lambda \kappa \sin(\phi - \tau s) \bs{e}_s \bs{e}_\phi - \Lambda \kappa \cos(\phi - \tau s) \bs{e}_r \bs{e}_s && \nonumber \\ + \Lambda \tau \bs{e}_r \bs{e}_\phi  + \Lambda \kappa \sin(\phi - \tau s) \bs{e}_\phi \bs{e}_s  -\Lambda \tau \bs{e}_\phi \bs{e}_r.  &&
\label{Appendix calculations: grad U bulk}
\end{eqnarray}
It is clear that $\bnabla \bs{U}_{\text{sym}} = \mathsfbi{0}$ in $\Omega \setminus \Omega_\delta$. For, $\bs{x} \in \Omega_\delta$, we have
\begin{eqnarray}
\bnabla \bs{U} =  - \frac{\Lambda (1 - r)}{\delta g^2} \frac{\partial g}{\partial s} \bs{e}_s \bs{e}_s + \frac{\Lambda \kappa (1 - r) \cos(\phi - \tau s)}{\delta g} \bs{e}_s \bs{e}_r && \nonumber \\
- \left[\frac{\Lambda \tau r (1 - r)}{\delta (1 - r \kappa \cos(\phi - \tau s)) g^2} \frac{\partial g}{\partial s} + \frac{\Lambda \kappa \sin(\phi - \tau s)(1-r)}{\delta g}\right] \bs{e}_s \bs{e}_\phi && \nonumber \\
-\left[\frac{\Lambda (1 - r \kappa \cos (\phi - \tau s))}{\delta g} + \frac{\Lambda \kappa \cos(\phi - \tau s) (1 - r)}{\delta g}\right]\bs{e}_r \bs{e}_s && \nonumber \\
+ \left[\frac{\Lambda \tau (1 - r)}{\delta g} - \frac{\Lambda \tau r}{\delta g}\right] \bs{e}_r \bs{e}_\phi - \frac{\Lambda \tau (1 - r)}{\delta g} \bs{e}_\phi \bs{e}_r - \frac{\Lambda \tau (1 - r)}{\delta g^2} \frac{\partial g}{\partial \phi} \bs{e}_\phi \bs{e}_\phi && \nonumber \\
+ \left[\frac{\Lambda \kappa \sin(\phi - \tau s) (1 - r)}{\delta g} - \frac{\Lambda (1 - r \kappa \cos(\phi - \tau s)) (1 - r) }{\delta r g^2} \frac{\partial g}{\partial \phi}\right]\bs{e}_\phi \bs{e}_s . && 
\label{Appendix calculations: grad U boundary layer}
\end{eqnarray}
 Given that $1-r$ is $O(\delta)$ in the boundary layer, calculation of $|\bnabla \bs{U}|^2$ up to leading order terms is
\begin{eqnarray}
|\bnabla \bs{U}|^2 = \frac{\Lambda^2 (1 - \kappa \cos (\phi - \tau s))^2}{\delta^2 g^2} + \frac{\Lambda^2 \tau^2}{\delta^2 g^2} + O(\delta^{-1}),
\label{Appendix calculations: grad U2 inter}
\end{eqnarray}
and
\begin{eqnarray}
|\bnabla \bs{U}|^2 h_s h_r h_\phi = \frac{\Lambda^2 (1 - \kappa \cos (\phi - \tau s))}{\delta^2 g^2} \left[(1 - \kappa \cos (\phi - \tau s))^2 + \tau^2 \right]  + O(\delta^{-1}).
\label{Appendix calculations: grad U2}
\end{eqnarray}
We use this result in (\ref{Bound on mean quantities: Q RHS 2nd}) for the calculation of $||\bnabla \bs{U}||^2$.
With the use of (\ref{Lower bound on the volume flow rate: constraint on g}), we see that the only two terms that are $O(\delta^{-1})$ in (\ref{Appendix calculations: grad U boundary layer}) are the terms involving $\bs{e}_r \bs{e}_s$ and $\bs{e}_r \bs{e}_\phi$. However, these terms do not contribute to the calculation of $\bs{U} \bcdot \bnabla \bs{U}$, as  they are multiplied with $U_r$ (the $r$ component of $\bs{U}$) which is zero. Therefore, $\bs{U} \bcdot \bnabla \bs{U}$ is $O(1)$ in $\Omega_\delta$. This result is useful in showing (\ref{Lower bound on the volume flow rate: U grad U}).

\subsection{Reason behind choice \ref{Lower bound on the volume flow rate: optimality condition}}
\label{Some useful calculations: optimality condition}
In the analysis done in the main text, if we had just assumed that $p_1$, $p_2$, and $p_3$ are constant functions but not necessarily equal, then a similar calculation would have led to
\begin{eqnarray}
g(s, \phi) = \sqrt{\frac{p_1 p_2 p_3}{p_3 (1-\kappa \cos(\phi - \tau s))^2 + p_1 \tau^2}} \qquad \qquad \qquad \qquad \nonumber \\
c_1(s, \phi) = \sqrt{\frac{p_1}{p_2} + \frac{p_1^2 \tau^2}{p_2 p_3 (1-\kappa \cos(\phi - \tau s))^2}}, \quad c_2(s, \phi) = \sqrt{\frac{p_3}{p_2} + \frac{p_3^2 (1-\kappa \cos(\phi - \tau s))^2}{p_1 p_2 \tau^2}}. \nonumber \\
\end{eqnarray}
With this choice, we could  have obtained the same bounds on the flow rate and the friction factor, namely
\begin{eqnarray}
Q \geq \sqrt{\frac{32 \upi^2}{27 I(\kappa, \tau)}}  +  O(\Rey^{-1}) \qquad \lambda \leq \lambda_b = \frac{27}{8} I(\kappa, \tau) +  O(\Rey^{-1}).
\end{eqnarray}
However, this time 
\begin{eqnarray}
I(\kappa, \tau) =  \frac{1}{2 \upi}\int_{0}^{2 \upi} (1-\kappa \cos \alpha) \left((1-\kappa \cos \alpha)^2 + \tau^2\right) \sqrt{\frac{M^2 (1-\kappa \cos \alpha)^2}{p_1^\prime} + \frac{M^2 \tau^2}{p_3^\prime}} \; d \alpha, \nonumber \\
\label{Appendix calculations: def I}
\end{eqnarray}
where
\begin{eqnarray}
M = \max\{p_1^\prime, 1, p_3^\prime\}, \quad p_1^\prime = \frac{p_1}{p_2}, \quad p_3^\prime = \frac{p_3}{p_2}.
\end{eqnarray}
To optimize the bound, we need to minimize $I(\kappa, \tau)$ and that clearly happens when 
\begin{eqnarray}
p_1^\prime = p_3^\prime = 1 \quad \implies p_1 = p_2 = p_3.
\end{eqnarray}

\bibliographystyle{jfm}
\bibliography{Reference}
\end{document}